\documentclass[aps,prd,floats,floatfix,nofootinbib,superscriptaddress]{revtex4-1}
\usepackage{graphicx,amsmath,amssymb,amsfonts,hyperref,mathdots,caption,subcaption, setspace, wrapfig,slashed,units}
\usepackage[usenames,dvipsnames,svgnames,table]{xcolor}
\usepackage[capitalize]{cleveref}

\newcommand{\e}[1]{\ensuremath{\mathrm{e}^{#1}}}
\newcommand{\mult}{\ensuremath{\Sigma}}
\newcommand{\co}{\ensuremath{^{*}}}
\newcommand{\ct}{\ensuremath{^{\dagger}}}
\newcommand{\zr}{\ensuremath{\zp{0,r}}}
\newcommand{\zi}{\ensuremath{\zp{0,i}}}
\newcommand{\zp}[1]{\ensuremath{\zeta^{#1}}}
\newcommand{\HkQ}[2]{\ensuremath{H_{#1}^{#2}}}
\newcommand{\aQ}{\ensuremath{\alpha_Q}}
\newcommand{\mr}{\ensuremath{m_{\zeta^{0,r}}}}
\newcommand{\mi}{\ensuremath{m_{\zeta^{0,i}}}}
\newcommand{\svSTD}{\ensuremath{\left<\sigma\beta\right>_{\rm STD}}}
\newcommand{\OZR}{\ensuremath{\frac{\Omega_{\zeta}}{\Omega_{\rm DM}}}}
\newcommand{\nOZR}{\ensuremath{\nicefrac{\Omega_\zeta}{\Omega_{\rm DM}}}}
\newcommand{\cm}[1]{\ensuremath{\widetilde{#1}}}
\newcommand{\Somm}{\ensuremath{\mathcal{S}}}
\newcommand{\sw}{s_W}
\newcommand{\ssw}{s_W^2}
\newcommand{\ccw}{c_W^2}
\newcommand{\cw}{c_W}
\newcommand{\alphaem}{\ensuremath{\alpha_{\rm EM}}}
\newcommand{\alphas}{\ensuremath{\alpha_{\rm s}}}

\newcolumntype{L}[1]{>{\raggedright\let\newline\\\arraybackslash\hspace{0pt}}m{#1}}
\newcolumntype{C}[1]{>{\centering\let\newline\\\arraybackslash\hspace{0pt}}m{#1}}
\newcolumntype{R}[1]{>{\raggedleft\let\newline\\\arraybackslash\hspace{0pt}}m{#1}}
\newcolumntype{D}[2]{>{\columncolor{#1}\centering\let\newline\\\arraybackslash\hspace{0pt}}m{#2}}

\begin{document}

\title{Large scalar multiplet dark matter in the high-mass region}

\author{Heather E.~Logan}
\email{logan@physics.carleton.ca}

\author{Terry Pilkington}
\email{tpilking@physics.carleton.ca}

\affiliation{Ottawa-Carleton Institute for Physics, Carleton University, 1125 Colonel By Drive, Ottawa, Ontario K1S 5B6, Canada}

\date{October 27, 2016}

\begin{abstract}
We study two models of scalar dark matter from ``large'' electroweak multiplets with isospin $5/2$ ($n=6$ members) and $7/2$ ($n=8$), whose scalar potentials preserve a $Z_2$ symmetry.  Because of large annihilation cross sections due to electroweak interactions, these scalars can constitute all the dark matter only for masses in the multi-TeV range.  For such high masses, Sommerfeld enhancement and co-annihilations play important roles in the dark matter relic abundance calculation, reducing the upper bound on the large multiplet's mass by almost a factor of two.  We determine the allowed parameter ranges including both of these effects and show that these models are as yet unconstrained by dark matter direct detection experiments, but will be probed by currently-running and proposed future experiments.  We also show that a Landau pole appears in these models at energy scales below $10^9$ GeV, indicating the presence of additional new physics below that scale.
\end{abstract}

\maketitle

\raggedbottom

\section{Introduction}

There is very strong evidence~\cite{Ade:2015xua,Vikram:2015leg} that the majority of matter in the Universe is in some form of dark matter (DM). The Standard Model of particle physics (SM) does not contain an appropriate DM candidate, which must satisfy the following criteria:
\begin{itemize}
\item{DM must interact gravitationally;}
\item{DM does not interact electromagnetically;}
\item{DM became non-relativistic at an early enough time; and}
\item{DM must be stable on cosmological timescales.}
\end{itemize}
Direct searches for DM have thus far produced no definite signal, only upper limits on the interaction cross section~\cite{Akerib:2015rjg}. The relic abundance of DM can be determined from the Cosmic Microwave Background (CMB) radiation, and is given by~\cite{Ade:2015xua}
\begin{equation}
\Omega_{\rm DM}\,h^2 = 0.1188\;,
\end{equation}
where $h$ is the dimensionless Hubble parameter defined by $H = 100 \cdot h~{\rm km/s/Mpc}$, and $\Omega_{\rm DM} = \rho_{\rm DM}/\rho_C$ is the fraction of the critical density, $\rho_C = 3H^2/(8\pi)$, in dark matter.

For weakly interacting massive particles (WIMPs) produced in the early Universe via standard thermal freeze-out (see, e.g., Ref.~\cite{Kolb:1990vq}), this gives a thermally-averaged cross section times relative velocity, $\beta$, as roughly~\cite{Steigman:2012nb}
\begin{equation}
\svSTD  = 3 \times 10^{-26}\;\mathrm{cm^3\,s^{-1}}\;.
\end{equation}
The requirement of satisfying this relic abundance, together with the need to evade current direct detection limits can place constraints on models that contain a DM candidate.

Extensions of the SM Higgs sector involving an additional ``inert'' scalar multiplet, the lightest state of which is stable and hence a possible DM candidate, have been well-studied in the isospin-singlet~\cite{singletrefs}, doublet~\cite{Deshpande:1977rw}, triplet~\cite{Araki:2011hm}, and quadruplet~\cite{AbdusSalam:2013eya} cases (for recent summaries of the experimental status of these models, see, e.g., Refs.~\cite{Cline:2013gha,Ayazi:2014tha,Belanger:2015kga,Queiroz:2014pra}). More recently, multiplets from larger representations of $SU(2)_L$ have been investigated in the context of dark matter~\cite{Cirelli:2005uq,Cirelli:2007xd,Cirelli:2009uv,Hambye:2009pw,Chowdhury:2015sla,Earl:2013jsa,Earl:2013fpa}, which examined both scalar and fermion multiplets.

In this paper, we expand upon the study in Ref.~\cite{Earl:2013fpa}. This study focused on models in which the SM is extended by a single large electroweak scalar multiplet, which is odd under an imposed global $Z_2$ symmetry. The multiplet carries hypercharge\footnote{We normalize $Y$ such that $Q = T^3 + Y/2$.} $Y_\mult = 1$, the same as the SM Higgs doublet, and has half-odd-integer weak isospin, $T = (n-1)/2$, where $n = 6$, $8$ counts the number of complex fields in the multiplet. Models with a larger complex scalar multiplet are disallowed by perturbative unitarity of the scattering of two scalars to two gauge bosons~\cite{Hally:2012pu}. As mentioned above, models with smaller multiplets have already been well-studied. Models in which the multiplet carries hypercharge $Y_\mult = 2T$, where the scalar potential preserves an accidental $U(1)$ symmetry, were studied in Ref.~\cite{Earl:2013jsa} and shown to be entirely excluded by dark matter direct detection constraints for $T > 2$.  Our objective in studying these two models is therefore to complete the analysis of \emph{all} DM models that extend the SM Higgs sector by a single (inert) scalar multiplet.

In Ref.~\cite{Earl:2013fpa}, it was shown that for DM candidate ($\zr$) masses around the weak scale, $80$~GeV~$\leq \mr \lesssim 1$~TeV, the DM candidate in these models can make up at most $1\%$ ($\nOZR \sim 0.01$) of the total DM content. However, since the fraction rises with DM candidate mass, we would na\"{i}vely expect there to be some mass where $\nOZR = 1$.  Above that mass, the model is excluded (assuming a standard thermal history) because the DM candidate would over-close the Universe. In the region of parameter space where this is expected to occur ($\mr \approx 20$~TeV), there are two additional effects which did not need to be considered in Ref.~\cite{Earl:2013fpa}. The first is co-annihilation: when the heavier scalars of the multiplet are close in mass to the lightest member, they will be present in roughly equal numbers in the thermal bath and will affect the freeze-out calculation. The second is Sommerfeld enhancement: in the non-relativistic limit of particle annihilation or scattering, the perturbative approach breaks down and we must consider the effects of an effective long-range force from the exchange of SM gauge bosons between the interacting particles.  We will show that these effects reduce the upper bound on the mass of the large multiplet by almost a factor of two.

We also study the renormalization group running of the quartic couplings in our models and determine the scale of the Landau pole.  It was shown in Ref.~\cite{Hamada:2015bra} that models with a large scalar multiplet develop a Landau pole at surprisingly low scales, even for vanishing quartic couplings at the weak scale.  We apply their results for the $n=6$ model and extend them to include the $n=8$ model, and show that the Landau pole appears at a scale at most 4 (2) orders of magnitude above the mass scale of the large multiplet in the $n=6$ (8) model.  Combining this with the upper bound on the DM mass to avoid over-closing the Universe, we show that the Landau pole must occur below $3 \times 10^8$~GeV in the $n=6$ model and below about $10^6$~GeV in the $n=8$ model, indicating that these models must be ultraviolet-completed well below the Planck scale.

This paper is organized as follows. In~\cref{sec:model}, we describe the model and set the notation. In~\cref{sec:naive} we calculate the relic abundance in the high DM-candidate mass region without the effects of co-annihilation and Sommerfeld enhancement. In~\cref{sec:coann}, we calculate the relic abundance including co-annihilating states and compare to the case of no co-annihilations. In~\cref{sec:sommerfeld}, we calculate the relic abundance including Sommerfeld enhancement of the single-particle annihilation and compare again to the original calculation. In~\cref{sec:coann_somm}, we calculate the relic abundance including both co-annihilation and Sommerfeld enhancement and compare to the other three cases. In~\cref{sec:landau_pole} we determine the scale of the Landau pole in the two models. In~\cref{sec:direct_detection} we describe the direct detection prospects of the models. We conclude in~\cref{sec:summary}. The generators for the larger representations of $SU(2)$, as well as the conjugation matrices, are given in~\cref{sec:matrices}. The relevant Feynman rules are provided in~\cref{sec:feynman_rules} (the full list is in Appendix B of Ref.~\cite{Earl:2013fpa}). Additionally, we provide the one-loop renormalization group equations for the scalar quartic couplings in our parameterization in~\cref{sec:RGEs}.~\cref{sec:iso_combos} lists the properly-normalized isospin combinations of pairs of large multiplets, which are used to construct the quartic terms in the scalar potential.

\section{Model description}\label{sec:model}

We consider two models that extend the SM through the addition of a single, large electroweak multiplet of complex scalars, $\mult$, which carries hypercharge $Y_\mult = 1$ and isospin $T = (n-1)/2$, where $n = 6$ (sextet) or $8$ (octet) is the size of the multiplet. In these models, the most general gauge-invariant scalar potential that preserves a $Z_2$ symmetry under which $\mult \to -\mult$ is given by
\begin{equation}\begin{aligned}\label{eq:conjugatepotential}
V(\Phi,\mult) &= m^2 \Phi^\dag \Phi + M^2 \mult^\dag \mult + \lambda_1 \big(\Phi^\dag \Phi\big)^2 + \lambda_2\Phi^\dag \Phi\, \mult^\dag \mult + \lambda_3 \Phi^\dag T_\Phi^a\Phi\, \mult^\dag T_\mult^a \mult\\
&\hspace*{3em} + \left[\lambda_4\,\widetilde{\Phi}^\dag T_\Phi^a \Phi\;\mult^\dag T_\mult^a\widetilde{\mult} + \mbox{h.c.}\right] + \mathcal{O}(\mult^4)\;,
\end{aligned}\end{equation}
where $\Phi$ is the SM SU(2)$_L$ doublet. Here $\widetilde{\Phi} = C\Phi\co$ and $\widetilde{\mult} = C\mult\co$ are the Higgs doublet and the large scalar multiplet in the conjugate representation, respectively. The conjugation matrix, $C$, is an antisymmetric $n \times n$ matrix equal to $i \sigma^2$ for the SU(2)$_L$ doublet. The $T_\Phi^a$ and $T_\mult^a$ matrices are the generators of $SU(2)_L$ in the doublet and $n$-plet representations, respectively. The matrices for $C$ and $T_{\mult}$ for $n=6$ and $8$ are given in~\cref{sec:matrices}.  The parameters $m^2$ and $\lambda_1$ are fixed in terms of the measured Higgs mass $m_h$ and the SM Higgs vacuum expectation value (vev) $v = (1/\sqrt{2} G_F)^{1/2} \simeq 246$~GeV by $\lambda_1 = m_h^2/2 v^2$ and $m^2 = -m_h^2/2$.  To ensure that the scalar potential has no alternative minima, a sufficient condition is that $M^2 > 0$ in the scalar potential. 

The term $\mult^\dag T^a\widetilde{\mult}$ (and its conjugate) can only be non-zero when $n$ is an even number (or, equivalently, $T$ is a half-odd-integer) which, combined with $n \leq 8$, restricts our models of interest to the two cases $n=6$ and $8$. For these two cases, the large multiplet is given in the electroweak basis by
\begin{equation}\begin{aligned}\label{eq:states}
\mult_{(n=6)} &= \left(\zp{+3},\,\zp{+2},\,\zp{+1},\,\zp{0},\, \zp{-1},\,\zp{-2}\right)^T\;,\\
\mult_{(n=8)} &= \left(\zp{+4},\,\zp{+3},\,\zp{+2},\,\zp{+1},\,\zp{0},\, \zp{-1},\,\zp{-2},\,\zp{-3}\right)^T\;.
\end{aligned}\end{equation}
Note that the conjugate of the charged state $\zeta^Q$ is written as $\zeta^{Q*}$, which is not the same as $\zeta^{-Q}$.

When the $\lambda_4$ term in Eq.~(\ref{eq:conjugatepotential}) vanishes, the Lagrangian preserves an accidental global $U(1)$ symmetry. Models with such a $U(1)$-symmetric potential have been studied in Ref.~\cite{Earl:2013jsa}. The inclusion of the $\lambda_4$ term has three effects. First, it breaks the would-be global $U$(1) symmetry down to a global $Z_2$ symmetry, under which $\mult \to -\mult$ and ${\rm SM} \to +{\rm SM}$. Second, the complex neutral component of $\mult$ is split into its real and imaginary parts, $\zr = \sqrt{2}~\mathrm{Re}~\zeta^0$ and $\zi = \sqrt{2}~\mathrm{Im}~\zeta^0$, with different masses. Finally, the states of $\mult$ with the same electric charge, $\zp{Q}$ and $\zp{-Q*}$, will mix to form mass eigenstates,
\begin{equation}\begin{aligned}\label{eq:mixed_states}
H_1^{Q} &= \cos \alpha_Q \, \zeta^{Q} + \sin \alpha_Q \, \zeta^{-Q*}\;,\\
H_2^{Q} &= -\sin \alpha_Q \, \zeta^{Q} + \cos \alpha_Q \, \zeta^{-Q*}\;,	
\end{aligned}\end{equation}
with $Q > 0$, $m_{H_1^Q} < m_{H_2^Q}$, and the mixing angle given by
\begin{equation}\begin{aligned}\label{eq:alphadef}
\tan\aQ &= (-1)^{\frac{n}{2} + Q +1} \frac{Q\lambda_3 - \sqrt{Q^2\lambda_3^2 + (n^2 - 4Q^2)\lambda_4^2}}{\lambda_4 \sqrt{n^2 - 4Q^2}}\\
&= (-1)^{\frac{n}{2} + Q} \frac{\lambda_4 \sqrt{n^2 - 4Q^2}}{Q\lambda_3 + \sqrt{Q^2\lambda_3^2 + (n^2 - 4Q^2)\lambda_4^2}}\;.
\end{aligned}\end{equation}
Since there is only one state with $Q = n/2$, the highest-charged state in the multiplet, it remains unmixed.

The masses of the physical states are given in terms of the mass of the neutral real particle, $m_{\zr}$, and the Lagrangian parameters $\lambda_3$ and $\lambda_4$, by~\cite{Earl:2013fpa},
\begin{equation}\begin{aligned}\label{eq:masses}
\mr^2 &= M^2 + \frac{1}{2}v^2\left[\lambda_2 + \frac{\lambda_3}{4} + \frac{n}{2}(-1)^{\frac{n}{2} + 1}\lambda_4\right]\;,\\
\mi^2 &= m_{\zr}^2 + \frac{n}{2}(-1)^{\frac{n}{2}}v^2\lambda_4\;,\\
m_{H_{1,2}^{+Q}}^2 &= m_{\zr}^2 + \frac{1}{4}v^2\left[n(-1)^{\frac{n}{2}}\lambda_4 \mp \sqrt{Q^2\lambda_3^2 + (n^2-4Q^2)\lambda_4^2}\right]\;,\\
m_{\zeta^{+\frac{n}{2}}}^2 &= m_{\zr}^2 - \frac{n}{8}v^2\left[\lambda_3 + 2 (-1)^{\frac{n}{2}+1}\lambda_4\right]\;, 
\end{aligned}\end{equation}
where the notation is such that the sign in $m_{H_{1,2}^Q}^2$ forces the relation $m_{H_1^Q} < m_{H_2^Q}$. The coupling of two $\zr$ to two Higgs bosons (the quantity in brackets in the definition of $\mr^2$) will be used as a scan parameter and is defined as
\begin{equation}\label{eq:Lambda_n_dfn}
\Lambda_n \equiv \lambda_2 + \frac{1}{4}\lambda_3 + \frac{n}{2}(-1)^{\frac{n}{2}+1}\lambda_4\;.
\end{equation}

For these models to contain a dark matter candidate, we require that the lightest (stable) member of the large multiplet be electrically neutral. This occurs only when $|\lambda_3| < 2|\lambda_4|$. We are then free to choose either $\zi$ or $\zr$ as the DM candidate. Without loss of generality, we choose the real part $\zr$ to be the lightest member of the large multiplet; this constrains the sign of $\lambda_4$ such that $\lambda_4 < 0$ for the sextet model and $\lambda_4 > 0$ for the octet.
The physical scalars arising from the large multiplet then always occur in the same mass ordering, given from lightest to heaviest by:
\begin{equation}\begin{aligned}\label{eq:masshierarchy}
&\zr,\, \HkQ{1}{\pm},\, \HkQ{1}{\pm\pm},\, \zp{\pm 3},\, \HkQ{2}{\pm\pm},\, \HkQ{2}{\pm},\, \zi, & (n = 6)\;,\\
&\zr,\, \HkQ{1}{\pm},\, \HkQ{1}{\pm\pm},\, \HkQ{1}{\pm 3},\, \zp{\pm 4},\, \HkQ{2}{\pm 3},\, \HkQ{2}{\pm\pm},\, \HkQ{2}{\pm},\, \zi, & (n = 8)\;.
\end{aligned}\end{equation}

In Ref.~\cite{Earl:2013fpa}, we showed that the parameter space can be constrained through perturbative unitarity of $2\to2$ scattering, electroweak precision measurements (the $STU$ observables), the rate for the decay of the Higgs boson to two photons as measured by the ATLAS and CMS experiments at the CERN Large Hadron Collider (LHC), and the absence of alternative minima in the scalar potential. For $\mr \leq 530$~GeV ($809$~GeV) in the $n=6$ ($8$) model, the constraints from $STU$ and $h \to \gamma\gamma$ limited $\zr$ to constitute less than $1\%$ of the total DM in the Universe. Furthermore, in Ref.~\cite{Logan:2015wba}, we showed that constraints arising from searches for new physics at the LHC were only sensitive to $\mr \lesssim 180$~GeV. In the mass region of interest in this paper ($\mr \gtrsim 1$~TeV), the only constraints on the parameter space come from the unitarity bounds on the quartic couplings $\lambda_i$, summarized in Table~\ref{tbl:unitarity_lambdas}, and the condition $M^2 >0$. Thus, we scan over the DM candidate mass and $\lambda_{2,3,4}$ that satisfy these constraints and calculate the relic abundance for the DM candidate.

\begin{table}
\begin{center}
{\def\arraystretch{1.33}
\begin{tabular}{C{2.5em}|C{4.5em}C{4.5em}C{4.5em}}
\hline\hline
$n$ & $|\lambda_2|^{\rm MAX}$ & $|\lambda_3|^{\rm MAX}$ & $|\lambda_4|^{\rm MAX}$\\
\hline
6 & 6.59 & 8.48 & 4.25\\
8 & 3.10 & 5.46 & 2.74\\
\hline\hline
\end{tabular}}
\end{center}
\caption{Upper bounds on the scalar quartic couplings from perturbative unitarity, from Ref.~\cite{Earl:2013fpa}. The values for $|\lambda_{2,3,4}|^{\rm MAX}$ were obtained using a coupled-channel analysis.}
\label{tbl:unitarity_lambdas}
\end{table}

\section{Relic abundance}

\subsection{Single-species calculation}
\label{sec:naive}

The relic abundance of $\zr$ is determined by its interactions in the early Universe. If we assume a standard thermal history---i.e., that the temperature was high enough at one time for $\zr$ to have been in thermal equilibrium, and that no late-decaying relics enhanced or diluted the $\zr$ density---then the relic density of $\zr$ at the present time can be computed from its annihilation rate in the early universe. For a generic relic, $X$, the density will be inversely proportional to the annihilation cross-section, $\Omega_X \propto \langle\sigma_X \beta \rangle^{-1}$~\cite{Steigman:2012nb}, where $\beta = v_{\rm rel}/c$ is the relative velocity of the two particles in the annihilation collision normalized to the speed of light and the brackets indicate an average over this velocity distribution at the time of freeze-out. Such an average is numerically necessary only if the annihilation cross section vanishes in the $\beta \to 0$ limit (which is not the case in our models). Because of this simple relationship, we can determine the fraction of the total dark matter that is made up of $X$ using the formula
\begin{equation}\label{eq:DM_frac}
\frac{\Omega_{X}}{\Omega_{\rm DM}} = \frac{\svSTD}{\left<\sigma \beta\left(X\;X \to {\rm SM}\;{\rm SM}\right)\right>}\;,
\end{equation}
where $\Omega_{\rm DM} h^2 = 0.1188$ is the current total dark matter relic abundance~\cite{Ade:2015xua}, and $\svSTD  = 3 \times 10^{-26}\;\mathrm{cm^3\,s^{-1}}$ is the ``standard'' annihilation cross section required to obtain this total dark matter relic abundance~\cite{Steigman:2012nb}.

For the large multiplet models, $X = \zr$, and the SM final states of interest are $W^{+}W^{-}$, $ZZ$, $hh$, and $f \bar{f}$ (via $s$-channel Higgs exchange). The DM fraction of~\cref{eq:DM_frac} is then given by
\begin{equation}\label{eq:DM_frac_naive}
\OZR = \frac{\svSTD}{\left<\sigma \beta\left(\zr\zr \to W^{+}W^{-},\, ZZ,\, hh,\, f\bar{f}\,\right)\right>}\;.
\end{equation}

The annihilation cross sections to gauge boson two-body final states were calculated in Ref.~\cite{Earl:2013fpa} and are given by
\begin{equation}\begin{aligned}\label{eq:dm_weak_scale_sigma_v_WpWm}
\sigma\beta (\zr\,\zr \to W^{+}\,W^{-}) = \frac{m_W^4}{8\pi\,v}\sqrt{1 - \frac{m_W^2}{\mr^2}}\Bigg[\frac{A_W^2}{\mr^2}\left(3 - 4\,\frac{\mr^2}{m_W^2} + 4\frac{\mr^4}{m_W^4}\right)&\\ + 2\,A_W\,B_W\,\left(1 - 3\,\frac{\mr^2}{m_W^2} + 2\,\frac{\mr^4}{m_W^4}\right)&\\ + B_W^2\,\mr^2\left(1 - \frac{\mr^2}{m_W^2}\right)^2\Bigg]&\;,
\end{aligned}\end{equation}
and
\begin{equation}\begin{aligned}\label{eq:dm_weak_scale_sigma_v_ZZ}
\sigma\beta (\zr\,\zr \to Z\,Z) = \frac{m_Z^4}{16\pi\,v}\sqrt{1 - \frac{m_Z^2}{\mr^2}}\Bigg[\frac{A_Z^2}{\mr^2}\left(3 - 4\,\frac{\mr^2}{m_Z^2} + 4\frac{\mr^4}{m_Z^4}\right)&\\ + 2\,A_Z\,B_Z\,\left(1 - 3\,\frac{\mr^2}{m_Z^2} + 2\,\frac{\mr^4}{m_Z^4}\right)&\\ + B_Z^2\,\mr^2\left(1 - \frac{\mr^2}{m_Z^2}\right)^2\Bigg]&\;,
\end{aligned}\end{equation}
where the coefficients are given by
\begin{equation}\begin{aligned}
A_Z &= 1 + \frac{\Lambda_n v^2}{4 \mr^2 - m_h^2}\;,\\
B_Z &= \frac{4}{m_Z^2 - \mr^2 - m_{\zeta^{0,i}}^2}\;,\\
A_W &= \frac{n^2 - 2}{2} + \frac{\Lambda_n v^2}{4 \mr^2 - m_h^2}\;,\\
B_W &= \frac{\left(n \cos\alpha_1 - \sqrt{n^2 - 4}\, \sin\alpha_1\right)^2}{m_W^2 - \mr^2 - m_{H_1^{+}}^2} + \frac{\left(-n \sin\alpha_1 - \sqrt{n^2 - 4}\, \cos\alpha_1\right)^2}{m_W^2 - \mr^2 - m_{H_2^{+}}^2}\;.
\end{aligned}\end{equation}
The mixing angles $\alpha_Q$ are given in~\cref{eq:alphadef} and the Higgs coupling $\Lambda_n$ in~\cref{eq:Lambda_n_dfn}. The annihilation cross sections to Higgs and fermion final states are given by~\cite{Earl:2013fpa}
\begin{equation}\label{eq:dm_weak_scale_sigma_v_hh}
\sigma\beta (\zr\,\zr \to h\,h) = \frac{\Lambda_n^2}{64\pi\,\mr^2}\sqrt{1 - \frac{m_h^2}{\mr^2}}\left[1 + \frac{3m_h^2}{4\mr^2 - m_h^2} - \frac{2\,v^2\,\Lambda_n}{2\mr^2 - m_h^2}\right]^2\;,
\end{equation}
and
\begin{equation}\label{eq:dm_weak_scale_sigma_v_ffbar}
\sigma\beta (\zr\,\zr \to f\,\bar{f}) = \frac{N_c}{4\pi}\left[1 - \frac{m_f^2}{\mr^2}\right]^{\frac{3}{2}}\,\frac{m_f^2\,\Lambda_n^2}{(4\mr^2 - m_h^2)^2}\;,
\end{equation}
where $N_c$ is the number of colours of the final-state fermions.

\begin{center}
\begin{figure}[t]
\begin{center}
\includegraphics[width=0.45\textwidth]{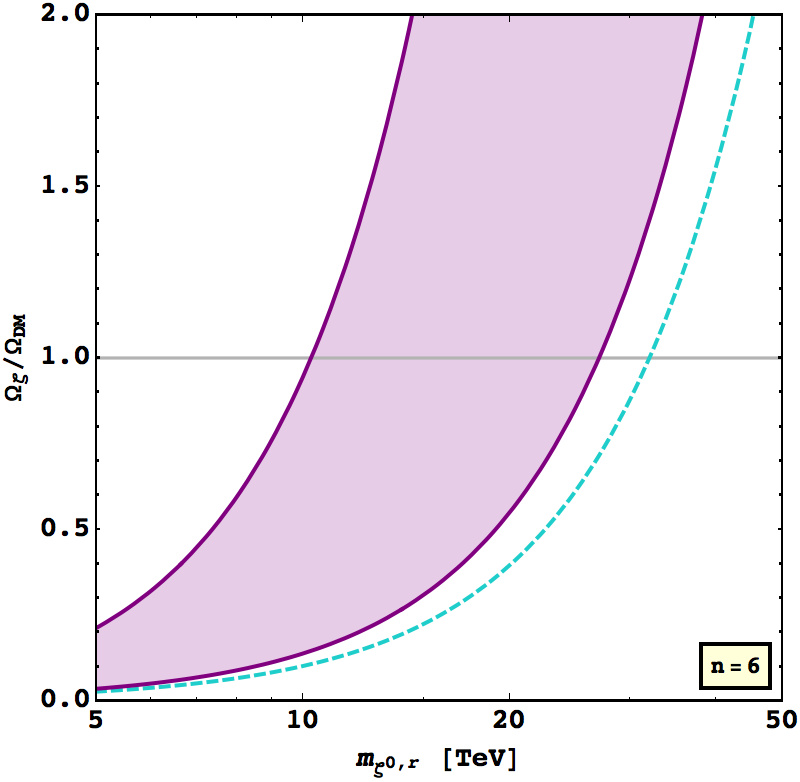}\hspace*{0.67em}
\includegraphics[width=0.45\textwidth]{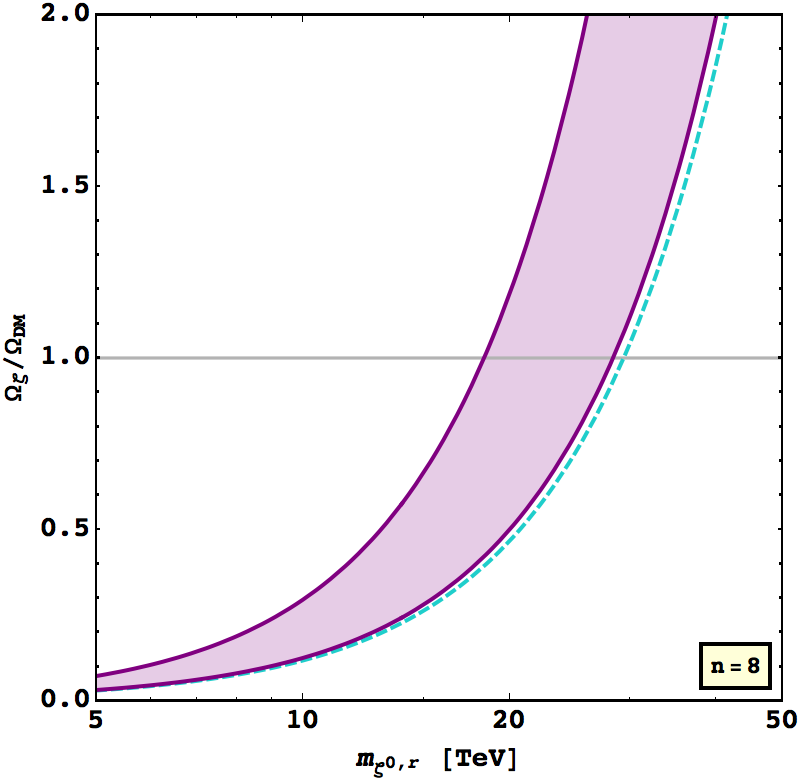}
\caption{The DM fraction $\nOZR$ as a function of $\mr$ for $n=6$ (left) and $n=8$ (right), computed solely from $\zeta^{0,r} \zeta^{0,r} \to$ SM SM. The allowed region for $\lambda_2 = \lambda_3 = 0$ is shaded purple. The dashed cyan lines show where $\nOZR = 1$ for $\Lambda_n = \Lambda_n^{\rm MAX}$ when $\lambda_{2,3} \neq 0$. As noted in the text, for the remainder of this paper we will set $\lambda_{2,3} = 0$ for simplicity, so that $\lambda_4 = \frac{2}{n}(-1)^{\frac{n}{2}}\Lambda_n$.}\label{fig:DM_frac_naive}
\end{center}
\end{figure}
\end{center}

As $\mr$ gets large, the cross section in each case falls like $\mr^{-2}$, which means that the DM fraction grows like $\mr^2$,
\begin{equation}\label{eq:DM_frac_high_m0}
\OZR \sim \mr^2 \svSTD\;.
\end{equation}
For $\mathcal{O}(1)$ quartic couplings, the pre-factor is also of order one. With $\nOZR = 1$, we find\footnote{$\sqrt{\frac{1}{\svSTD}} = 19.7$~TeV.} that $\mr \sim 20$~TeV. From Refs.~\cite{Earl:2013fpa,Logan:2015wba}, we know that the only constraints on this region of parameter space come from perturbative unitarity (numerical values given in~\cref{tbl:unitarity_lambdas}) and stability of the potential ($M^2 > 0$). To simplify later calculations, we set $\lambda_{2,3} = 0$ and scan over $\Lambda_n = \frac{n}{2}|\lambda_4|$. We calculate the relic abundance and plot the result as the shaded regions in~\cref{fig:DM_frac_naive} (the left panel for $n=6$ and the right panel for $n=8$). This shaded region is that which is allowed by perturbative unitarity and $M^2>0$. From this, we find that $\nOZR = 1$ for $10.1$~TeV~$\leq \mr \leq 27.0$~TeV in the $n=6$ model, and $18.4$~TeV~$\leq \mr \leq 28.4$~TeV in the $n=8$ model.

To ensure that this is a valid simplification, we also scan over $\lambda_{2,3} \neq 0$. The lower-bound (left-most solid purple curve in~\cref{fig:DM_frac_naive}) does not change, as this is where $\lambda_2 = \lambda_3 = \lambda_4 = 0$. The upper-bound is shown as the dashed cyan curve, which gives the mass range for $\nOZR = 1$ as $10.1$~TeV~$\leq \mr \leq 32.0$~TeV in the $n=6$ model, and $18.4$~TeV~$\leq \mr \leq 29.3$~TeV in the $n=8$ model. This difference introduces a small uncertainty in the upper bound of the mass range (particularly in the $n=6$ case) when we use the $\lambda_{2,3} = 0$ approximation.

\subsection{Co-annihilations}\label{sec:coann}

The masses of all the $Z_2$-odd scalars in $\mult$ will fall between two values: $\mr$ and $\mi$, as in~\cref{eq:masshierarchy}. The unitarity bounds, however, do not depend on the overall mass scale of $\mult$. This means that the overall mass splitting will become squeezed as $\mr$ increases, according to
\begin{equation}
\Delta m_0 \equiv \mi - \mr = \sqrt{\mr^2 + \frac{n}{2}\,(-1)^{\frac{n}{2}}\,v^2\,\lambda_4} - \mr \approx \frac{n}{4}|\lambda_4|\,\frac{v^2}{\mr} + \mathcal{O}\left(\frac{v^4}{\mr^3}\right)\;.
\end{equation}

In Ref.~\cite{Earl:2013fpa}, we made the assumption that all of the heavier states of $\mult$ had decayed so that only $\zr$ remained at freeze-out. When the mass splitting is compressed, as in the current case ($\Delta m_0 < m_W$ for $\mr \gtrsim 5$~TeV), the ``heavier" states of $\mult$ will still be present in the thermal bath during freeze-out. These co-annihilating states will affect the relic abundance of $\zr$. In what follows we make the approximation that all members of $\mult$ are degenerate as far as the equilibrium number density is concerned.\footnote{As we will see, because freeze-out happens at temperatures above the electroweak phase transition, this approximation will become exact.}

The DM fraction from~\cref{eq:DM_frac} becomes
\begin{equation}\label{eq:DM_frac_coann}
\OZR = \frac{\svSTD}{\frac{1}{(2n)^2}\sum_Q\,\left<\sigma\beta({\{\mult\,\mult\}}^Q \to {\{{\rm SM\,SM}\}}^Q)\right>}\;,
\end{equation}
where the sum is over the appropriate SM final-state charges, $Q = 0,\,\pm1,\,\pm2$, and the factor of $\nicefrac{1}{(2n)^2}$ in the denominator is the average over initial species and accounts for particles not meeting the ``right" partner to annihilate in the early Universe. The required charge combinations from $\mult$ are, for $n=6$,
\begin{equation}\begin{aligned}
&{\{\mult\,\mult\}}^0 &\in&& &\left\{\zr\zr,\,\zr\zi,\,\zi\zi,\,\HkQ{k}{+q}\HkQ{\ell}{-q},\,\zp{+3}\zp{-3}\right\}\;,\\
&{\{\mult\,\mult\}}^{\pm1} &\in&& &\left\{\zr\HkQ{k}{\pm1},\,\zi\HkQ{k}{\pm1},\,\HkQ{k}{\pm q}\HkQ{\ell}{\mp(q-1)},\,\zp{\pm3}\HkQ{k}{\mp2}\right\}\;,\\
&{\{\mult\,\mult\}}^{\pm2} &\in&& &\left\{\zr\HkQ{k}{\pm2},\,\zi\HkQ{k}{\pm2},\,\HkQ{k}{\pm1}\HkQ{\ell}{\pm1},\,\zp{\pm3}\HkQ{k}{\mp1}\right\}\;,
\end{aligned}\end{equation}
and for $n=8$,
\begin{equation}\begin{aligned}
&{\{\mult\,\mult\}}^0 &\in&& &\left\{\zr\zr,\,\zr\zi,\,\zi\zi,\,\HkQ{k}{+q}\HkQ{\ell}{-q},\,\zp{+4}\zp{-4}\right\}\;,\\
&{\{\mult\,\mult\}}^{\pm1} &\in&& &\left\{\zr\HkQ{k}{\pm1},\,\zi\HkQ{k}{\pm1},\,\HkQ{k}{\pm q}\HkQ{\ell}{\mp(q-1)},\,\zp{\pm4}\HkQ{k}{\mp3}\right\}\;,\\
&{\{\mult\,\mult\}}^{\pm2} &\in&& &\left\{\zr\HkQ{k}{\pm2},\,\zi\HkQ{k}{\pm2},\,\HkQ{k}{\pm1}\HkQ{\ell}{\pm1},\,\HkQ{k}{\pm3}\HkQ{\ell}{\mp1},\,\zp{\pm4}\HkQ{k}{\mp2}\right\}\;,
\end{aligned}\end{equation}
where $k,\ell = 1,2$ and $q = 1,\,\dots,\,\nicefrac{n}{2}-1$. The SM combinations are
\begin{equation}\begin{aligned}
&{\{{\rm SM\,SM}\}}^0 &\in&& &\left\{W^{+}W^{-},\,ZZ,\,Z\gamma,\,\gamma\gamma,\,hh,\,h\gamma,\,hZ,\,f\bar{f}\right\}\;,\\
&{\{{\rm SM\,SM}\}}^{\pm1} &\in&& &\left\{W^{\pm}Z,\,W^{\pm}\gamma,\,W^{\pm}h,\,ff'\right\}\;,\\
&{\{{\rm SM\,SM}\}}^{\pm2} &\in&& &\left\{W^{\pm}W^{\pm}\right\}\;.
\end{aligned}\end{equation}

\begin{center}
\begin{figure}[t]
\begin{center}
\includegraphics[width=0.45\textwidth]{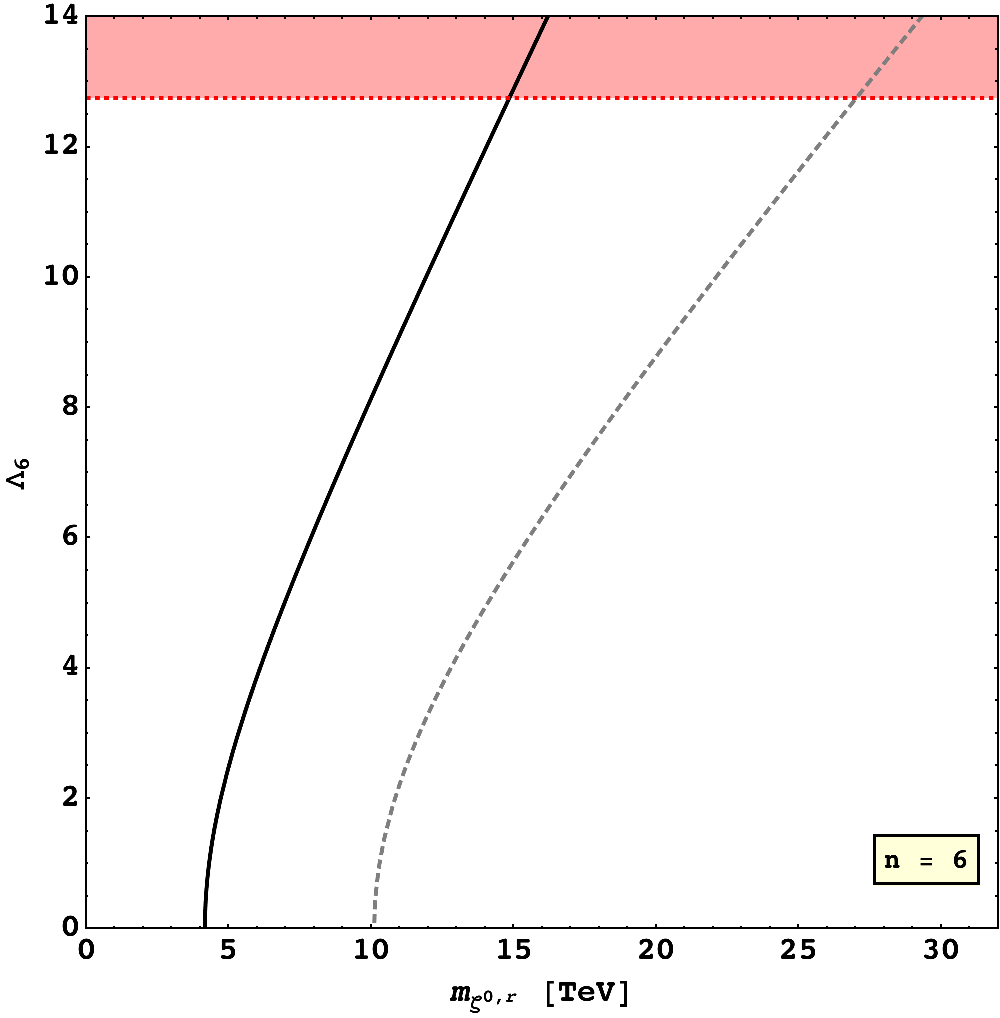}\hspace*{1em}
\includegraphics[width=0.45\textwidth]{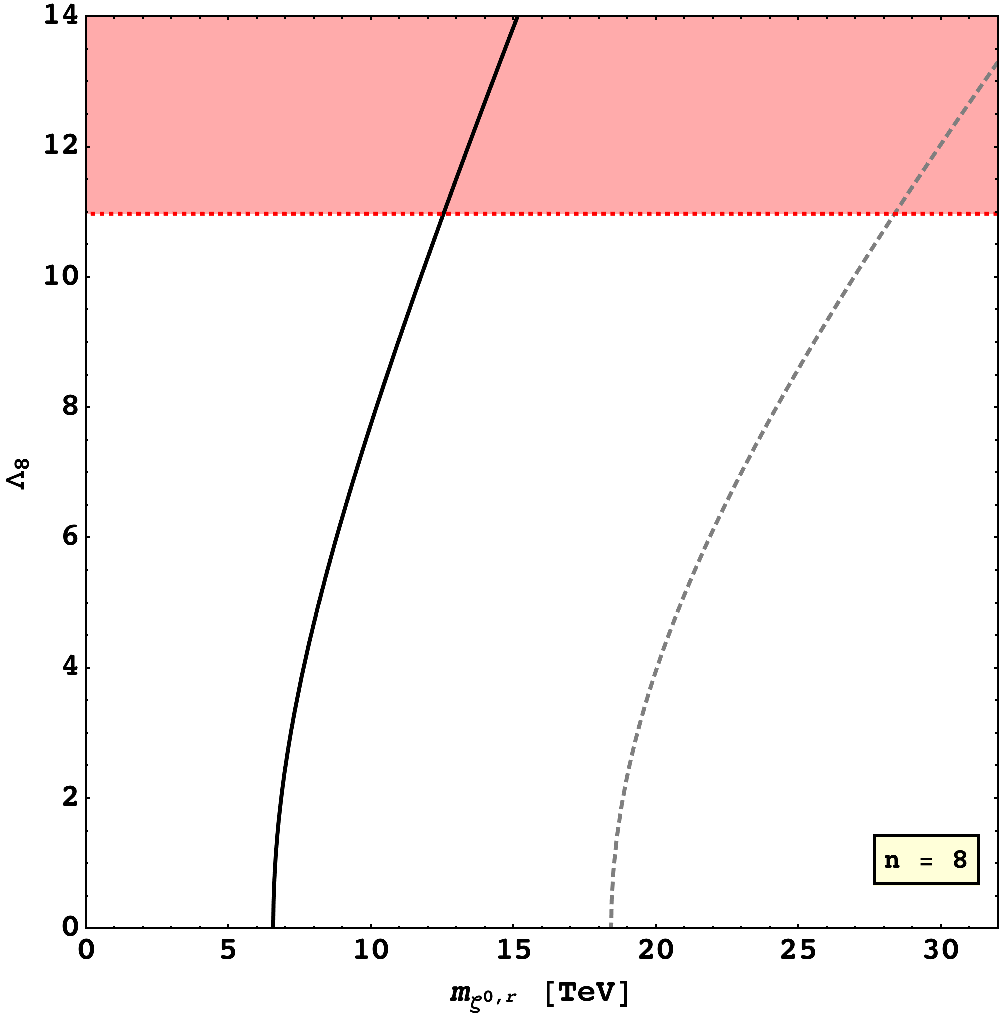}
\caption{Parameter values for which $\nOZR = 1$ for co-annihilation (black solid) and single-species annihilation (dashed grey), as a function of both $\mr$ and $\Lambda_n$. The region above the horizontal dotted red line is excluded by the unitarity bound, $\Lambda_n \leq \Lambda_n^{\rm MAX}$.  We set $\lambda_2 = \lambda_3 = 0$.}\label{fig:DM_frac_coann}
\end{center}
\end{figure}
\end{center}

We set $\lambda_2 = \lambda_3 = 0$ and $\lambda_4 = \frac{2}{n}(-1)^{\frac{n}{2}}\Lambda_n$, and scan over $\mr$ and $\Lambda_n$ to determine where $\nOZR = 1$. We plot the results in~\cref{fig:DM_frac_coann} where the solid black curve corresponds to $\nOZR = 1$ for the co-annihilating case,~\cref{eq:DM_frac_coann}, and the dashed grey curve corresponds to $\nOZR = 1$ for the single particle annihilation case,~\cref{eq:DM_frac_naive}. The region above the horizontal dotted red line is ruled out by the unitarity bound from~\cref{tbl:unitarity_lambdas}.

When we compare the expressions for the DM fraction,~\cref{eq:DM_frac_naive,eq:DM_frac_coann}, we would expect that the denominator in the co-annihilation case would numerically be much larger than in the na\"{i}ve single particle annihilation case---more contributions to the cross section will necessarily increase the total cross section, and decrease the DM fraction.  However, being that there are more members of the multiplet present in the thermal bath, it is more likely that any two particles that meet will not be able to annihilate (e.g., there is no two-body SM final state that could accommodate the situation where $\zr$ meets $\zp{+4}$). This is taken into account by averaging over the number of species $2n$ in each of the two incoming ``beams''. This factor in the denominator substantially decreases the total cross section over the entire mass range, increasing the DM fraction. In both cases, the relic density is pushed up so that the mass range required for $\nOZR = 1$ is lower. In the case where co-annihilations are present, we find that $\nOZR = 1$ corresponds to $4.2$~TeV~$\leq \mr \leq 14.9$~TeV in $n=6$ and $6.5$~TeV~$\leq \mr \leq 12.5$~TeV in $n=8$.

\subsection{Sommerfeld enhancement}\label{sec:sommerfeld}

Members of the large multiplet will annihilate via gauge and Higgs bosons. At non-relativistic speeds, and at high $\mr \gg m_{W,Z,h}$, radiative corrections to the annihilation cross section will be important. In that case, the gauge and Higgs bosons mediate an effective long-range force between the annihilating particles. This effect is known as the Sommerfeld enhancement~\cite{Sommerfeld:1931aa}. Its importance for DM annihilation was first described in Ref.~\cite{Hisano:2003ec}. For pedagogical overviews see, e.g., Refs.~\cite{ArkaniHamed:2008qn,Slatyer:2009vg,Cirelli:2007xd}. In this section, we consider only the single-particle annihilation case ($\zr\,\zr \to {\rm SM}\,{\rm SM}$). In the following section we compute the Sommerfeld enhancement including co-annihilating states.

Consider a particle moving non-relativistically through space. The wave-function that describes the particle is a solution to the time-independent Schr\"{o}dinger equation,
\begin{equation}\label{eq:simple_schrodinger}
-\frac{1}{2m}\,\nabla^2\psi^{(0)}(\vec{r}\,) = E\,\psi^{(0)}(\vec{r}\,)\;.
\end{equation}
The probability density at the origin will be $|\psi^{(0)}(0)|^2$.

If we now introduce a central potential, $V(r)$, which may be attractive or repulsive, then we will modify the wave function at the origin.  For $r$ such that $|E| \lesssim |V|$, the potential will distort the wave function. If we consider the case where we have plane waves coming in, which scatter from the potential $V(r)$, and spherical waves are seen exiting at large $r$, then we have the asymptotic solution
\begin{equation}
\psi(\vec{r}\,) \stackrel{r \to \infty}{\longrightarrow} \e{i\,k\,z} + f(\theta)\,\frac{\e{i\,k\,r}}{r}\;.
\end{equation}
The probability density at the origin will be $|\psi(0)|^2$.

The cross section for a short-distance annihilation process will be proportional to the square of the amplitude of the wave function at the origin. If we compare the cross section to its unperturbed value,
\begin{equation}\label{eq:somm_def}
\frac{\sigma}{\sigma_0} = \frac{|\psi(0)|^2}{|\psi^{(0)}(0)|^2} \equiv \Somm\;,
\end{equation}
then we may define $\Somm$ as the Sommerfeld factor~\cite{Sommerfeld:1931aa}.

The exchange of SM particles between the DM acts as a long-range force, which affects the annihilation cross section as in~\cref{eq:somm_def}. Because it affects the cross section only as a multiplicative factor, we may factorize the calculation into a long-range (Sommerfeld) part and a short-range (annihilation) part. For Coulomb-like scattering,
\begin{equation}
V_{\rm Coul.} = -\eta\frac{\alpha}{r}\;,
\end{equation}
where $\eta = +1$ ($-1$) for an attractive (repulsive) Coulomb potential and $\alpha$ is the coupling strength.  Given the relative velocity, $\beta$, the Sommerfeld factor is~\cite{Harris:1957zza}
\begin{equation}\label{eq:dm_somm_coulomb}
\Somm_{\rm Coul.} = \eta\frac{2\pi}{\epsilon_\beta}\,\frac{1}{1 - \e{-\eta\frac{2\pi}{\epsilon_\beta}}}\;,
\end{equation}
where $\epsilon_\beta \equiv \nicefrac{\beta}{\alpha}$.

For the case of a Yukawa-like potential, the potential is
\begin{equation}
V_{\rm Yuk.} = \pm\frac{\alpha}{r}\,\e{-m\,r}\;,
\end{equation}
where, as in the Coulomb case, $-$ corresponds to an attractive potential, $+$ corresponds to a repulsive potential, $\alpha$ is the coupling strength, and now $m$ is the mass of the exchanged particle. Unfortunately, this does not have a nice, analytic solution like $V_{\rm Coul.}$, and we would need to determine $\Somm$ numerically. Fortunately, though, there is a similar potential, the Hulth\'{e}n potential~\cite{Hulthen:1942aa}, which exhibits the same behaviour in both the small-$r$ and large-$r$ limits. It was shown in Ref.~\cite{Feng:2010zp} that the Hulth\'{e}n potential reproduces the numerical results for the Sommerfeld factor due to the Yukawa potential to better than $10\%$, and accurately describes the resonant behaviour which will be discussed below. More importantly, the time-independent Schr\"{o}dinger equation with the Hulth\'{e}n potential does have an analytic solution. The Hulth\'{e}n potential is given by
\begin{equation}
V_{\rm Hulth\acute{e}n} = \pm\alpha\,\omega\,\frac{\e{-\omega\,r}}{1 - \e{-\omega\,r}}\;,
\end{equation}
where $\omega \equiv \nicefrac{6m}{\pi^2}$. The Sommerfeld enhancement factor for the Hulth\'{e}n potential is given by~\cite{Slatyer:2009vg} (see also Ref.~\cite{Cassel:2009wt})
\begin{equation}\label{eq:dm_multi_TeV_somm_hulthen}
\Somm_{\rm Hulth\acute{e}n} = \frac{\pi}{\epsilon_\beta}\,\frac{\sinh\left(\frac{2\pi\,\epsilon_\beta}{\epsilon_\omega}\right)}{\cosh\left(\frac{2\pi\epsilon_\beta}{\epsilon_\omega}\right) - \cos\left(2\pi\sqrt{\frac{1}{\epsilon_\omega} - \frac{\epsilon_\beta^2}{\epsilon_\omega^2}}\right)}\;,
\end{equation}
where $\epsilon_\beta \equiv \nicefrac{\beta}{\alpha}$, $\epsilon_\omega \equiv \nicefrac{\omega}{\alpha\,M_{\rm DM}}$, and $M_{\rm DM}$ is the mass of the scattering particles.

The calculation of the potential, $V(r)$, is done using the Born approximation (see, e.g., Chapter 4 of Ref.~\cite{Peskin:1995ev}), where
\begin{equation}
V(r) = \int \frac{d^3 q}{(2\pi)^3}\,\widetilde{V}(\vec{q}\,)\,e^{i \vec{q}\cdot\vec{r}}\;,
\end{equation}
where $\widetilde{V}(\vec{q}\,)$ is minus the tree-level matrix element in the soft scattering limit and $\vec{q}$ is the $t$-channel momentum transfer. The scattering may proceed in general via the exchange of $\gamma$, $W^{\pm}$, $Z$, or $h$. Kinematic factors are suppressed because we are working in the low momentum transfer limit.

The corresponding potentials are given by
\begin{equation}\begin{aligned}
V_{\gamma}(r) &= -\left[\alpha_{\rm EM}\, C_{s_1 s_1 \gamma} C_{s_2 s_2 \gamma}\right]\frac{1}{r} &&\equiv -g_\gamma \frac{1}{r}\;,\\
V_{W}(r) &= -\left[\alpha_{\rm EM}\, C_{s_1 s_4 W} C_{s_2 s_3 W}^{*}\right] \frac{\e{-m_W r}}{r} &&\equiv -g_W \frac{\e{-m_W r}}{r}\;,\\
V_{Z}(r) &= -\left[\alpha_{\rm EM}\, C_{s_1 s_4 Z} C_{s_2 s_3 Z}^{*}\right] \frac{\e{-m_Z r}}{r} &&\equiv -g_Z \frac{\e{-m_Z r}}{r}\;,\\
V_{h}(r) &= -\left[\frac{C_{s_1 s_1 h} C_{s_2 s_2 h}}{4\pi}\right] \frac{\e{-m_h r}}{r} &&\equiv -g_h \frac{\e{-m_h r}}{r}\;,
\end{aligned}\end{equation}
where the couplings $C_{ijk}$ are given in~\cref{sec:feynman_rules} and we define $g_{\gamma,\,W,\,Z,\,h}$ as the couplings in the brackets. Notice on the far right-hand side, each of the last three potentials has a Yukawa form, and so we may use the Hulth\'{e}n approximation.

In the case of $\zr\zr \to {\rm SM}\,{\rm SM}$, the only relevant potential is $V_h(r)$, as the others are all zero for $\zr$ scattering. At temperatures above the electroweak phase transition, which would normally be the case in the multi-TeV mass region, the Higgs vev would be zero (see the next section). In that case, the $\zr\zr h$ coupling would also be zero, and there would be no Sommerfeld enhancement from Higgs exchange. However, we would like to examine the form of the effects of Sommerfeld enhancement in this simpler case before moving on to combining Sommerfeld enhancement with co-annihilation, so we retain $v \neq 0$ for now. Then the Sommerfeld enhancement factor $\Somm_h$ from Higgs exchange is given by~\cref{eq:dm_multi_TeV_somm_hulthen} with
\begin{equation}\begin{aligned}
\epsilon_\beta &= \frac{\beta}{\alpha_\zeta}\;, &&&
\epsilon_\omega &= \frac{\pi^2}{6}\,\frac{m_h}{\alpha_\zeta\,\mr}\;, &&&
&\mbox{and}&&&
\alpha_\zeta &= \frac{\Lambda_n^2}{4\pi}\;.
\end{aligned}\end{equation}

\begin{center}
\begin{figure}[t]
\begin{center}
\includegraphics[width=0.45\textwidth]{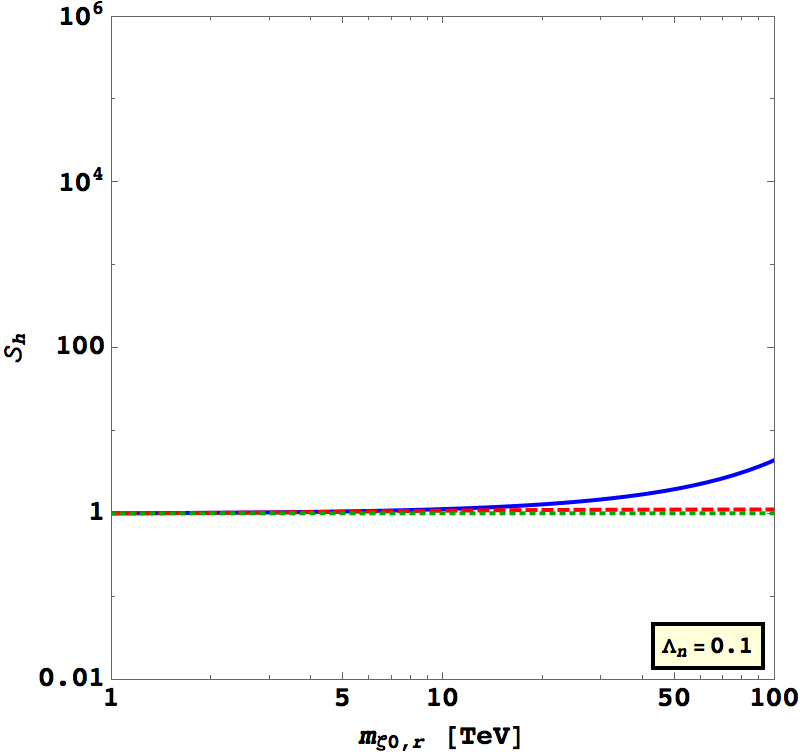}\hspace*{0.67em}
\includegraphics[width=0.45\textwidth]{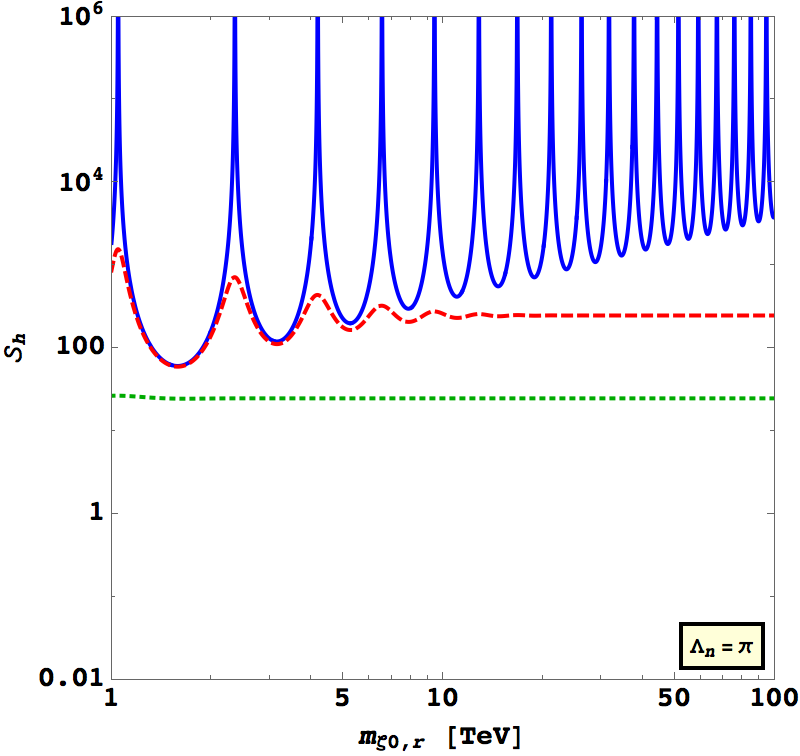}
\caption{Sommerfeld enhancement due to the Hulth\'{e}n potential for $\Lambda_n = 0.1$ (left plot) and $\Lambda_n = \pi$ (right plot). The curves correspond to different values of $\beta = \nicefrac{v_{\rm rel}}{c}$. The solid blue curve corresponds to $\beta = 10^{-5}$, the dashed red to $\beta = 10^{-2}$, and the dotted green to $\beta = 10^{-1}$.}\label{fig:dm_multi_TeV_somm_enhance_hulth}
\end{center}
\end{figure}
\end{center}

We plot in~\cref{fig:dm_multi_TeV_somm_enhance_hulth} the Sommerfeld enhancement factor for $\zr\,\zr$ via the exchange of a Higgs boson using the Hulth\'{e}n potential, where $\omega = \nicefrac{6m_h}{\pi^2}$, $M_{\rm DM} = \mr$, and $\alpha = \nicefrac{\Lambda_{n}^2}{4\pi}$. Notice the structure present in these plots that is not present in the Coulomb case. As $\beta \to 0$, the energy of the incident particles is near zero. The number of bound states in the Coulomb case is infinite, and so changing the parameters of the potential (in this case, $\alpha$) will have no effect on the number of bound states. On the other hand, in the Yukawa (or Hulth\'{e}n) potential, there are a finite number of bound states. If the potential is modified (in this case, either by changing $\epsilon_\beta$ or $\epsilon_\omega$), the number of bound states may change. That is, changes in these parameters will bring bound states close to $E = 0$. If the energy of the incident particle (e.g., $E = +\delta$ for some small $\delta$) is close to that of a bound state near $E=0$, then the interaction will be resonantly enhanced, leading to the spikes in the right-hand plot of~\cref{fig:dm_multi_TeV_somm_enhance_hulth}, where the coupling is large and the velocity is small.

\begin{center}
\begin{figure}[t]
\begin{center}
\includegraphics[width=0.45\textwidth]{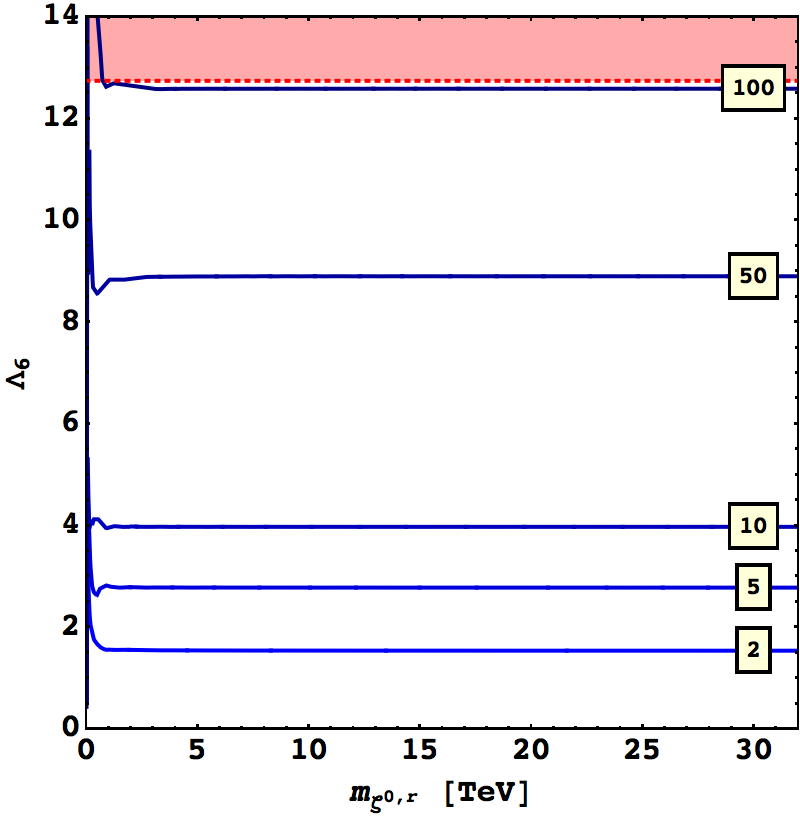}\hspace*{0.67em}
\includegraphics[width=0.45\textwidth]{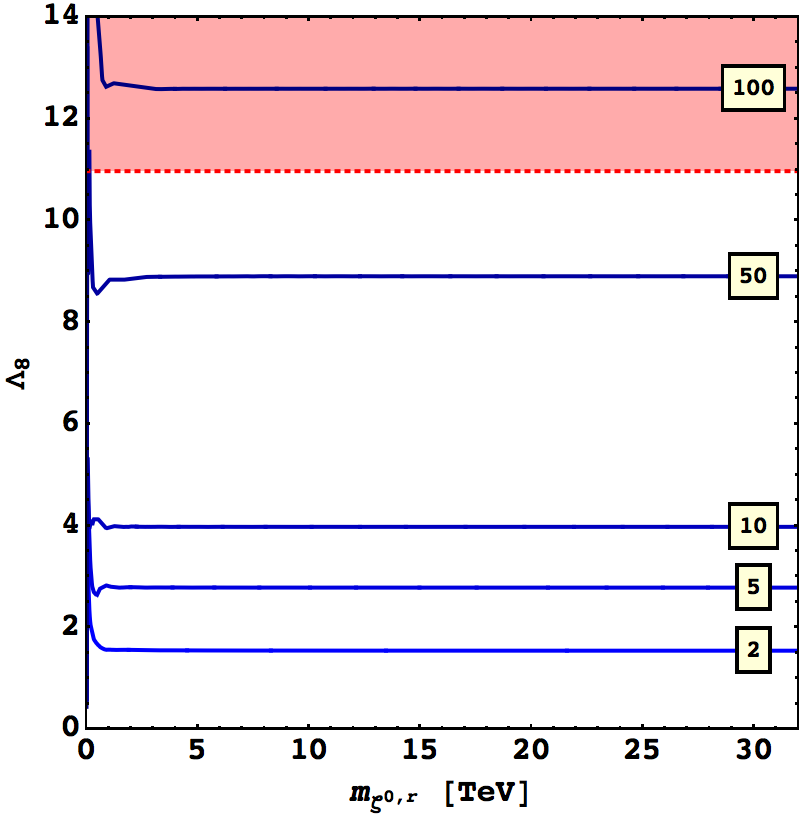}
\caption{Thermally-averaged Sommerfeld enhancement factor from the Hulth\'{e}n potential as a function of $\mr$ and $\Lambda_n$. Contours are labelled with the value of $\langle \Somm_h \rangle$ for $x = 20$. Since $\langle \Somm_h \rangle$ does not depend on the size of the multiplet, the contours in the $n = 6$ (left plot) and $n = 8$ (right plot) models are the same. Except for numerical instability as $\mr \to 0$ (i.e., $\mr \approx m_{W,Z,h}$, so that the effective long-range potential vanishes), the Sommerfeld factor does not vary with $\mr$, but does increase with increasing $\Lambda_n$. The region above the horizontal dotted red line is excluded by the unitarity bound, $\Lambda_n \leq \Lambda_n^{\rm MAX}$.}\label{fig:dm_multi_TeV_therm_somm}
\end{center}
\end{figure}
\end{center}

Returning to the annihilation cross section,~\cref{eq:somm_def}, we will have $\sigma\beta = \Somm_h\cdot\sigma_0\beta$, where $\sigma_0\beta$ is the same sum of the cross section times velocity as in~\cref{eq:DM_frac_naive}. Notice that $\Somm_h$ is a function of $\beta$, and so we will have to thermally-average its contribution,
\begin{equation}
\left<\sigma\beta\right> = \frac{x^{\frac{3}{2}}}{4\pi}\int\,\Somm_h\,\sigma_0\beta\,\e{-\frac{x\,\beta^2}{4}}\,\beta^2\,d\beta\; ,
\end{equation}
where $x = m_{\zeta^{0,r}}/T$ and $T$ is the temperature.
As before, in the case of s-wave annihilation, the cross section times velocity $\sigma_0 \beta$ is not a function of $\beta$, so it may be taken outside of the integral, and we are left with
\begin{equation}
\left<\sigma\beta\right> = \sigma_0\beta\,\left<\Somm_h \right>\;,
\end{equation}
where
\begin{equation}
\left<\Somm_h\right> \equiv \frac{x^{\frac{3}{2}}}{4\pi}\int\,\Somm_h\,\e{-\frac{x\,\beta^2}{4}}\,\beta^2\,d\beta\;.
\end{equation}
We evaluate this numerically. The effect of Sommerfeld enhancement in the early Universe is relatively small, due to the high-$\beta$ tail of the velocity distribution---recall that Sommerfeld enhancement is strongest when $\beta$ is small. We plot $\left<\Somm_h\right>$ as a function of $\mr$ and $\Lambda_n$ for the standard freeze-out value of $x = 20$ in~\cref{fig:dm_multi_TeV_therm_somm}. The contours are labelled with the value of $\left<\Somm_h\right>$. The thermally-averaged Sommerfeld enhancement factor varies with $\Lambda_n$, but not with $\mr$. Compare this to~\cref{fig:dm_multi_TeV_somm_enhance_hulth}, where $\Somm_h$ was not thermally-averaged and exhibits resonance behaviour.

\begin{center}
\begin{figure}[t]
\begin{center}
\includegraphics[width=0.45\textwidth]{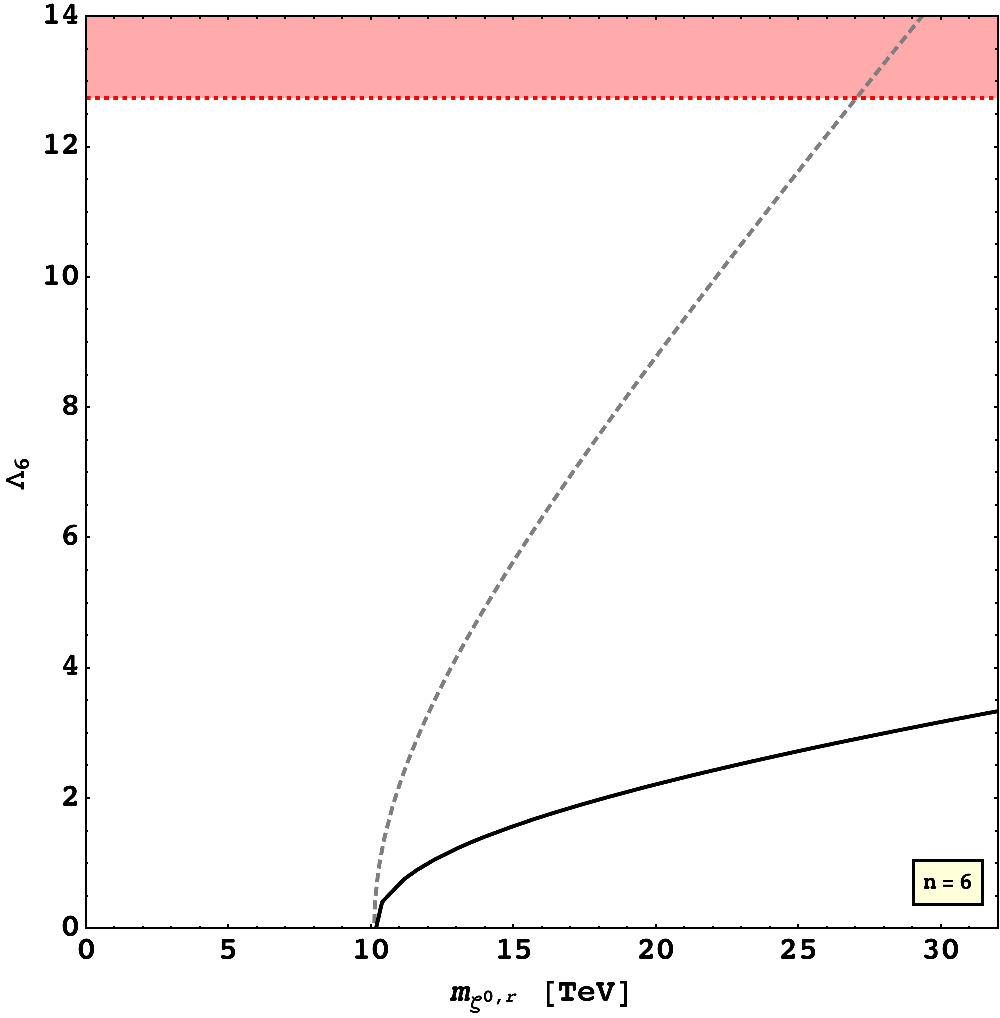}\hspace*{0.67em}
\includegraphics[width=0.45\textwidth]{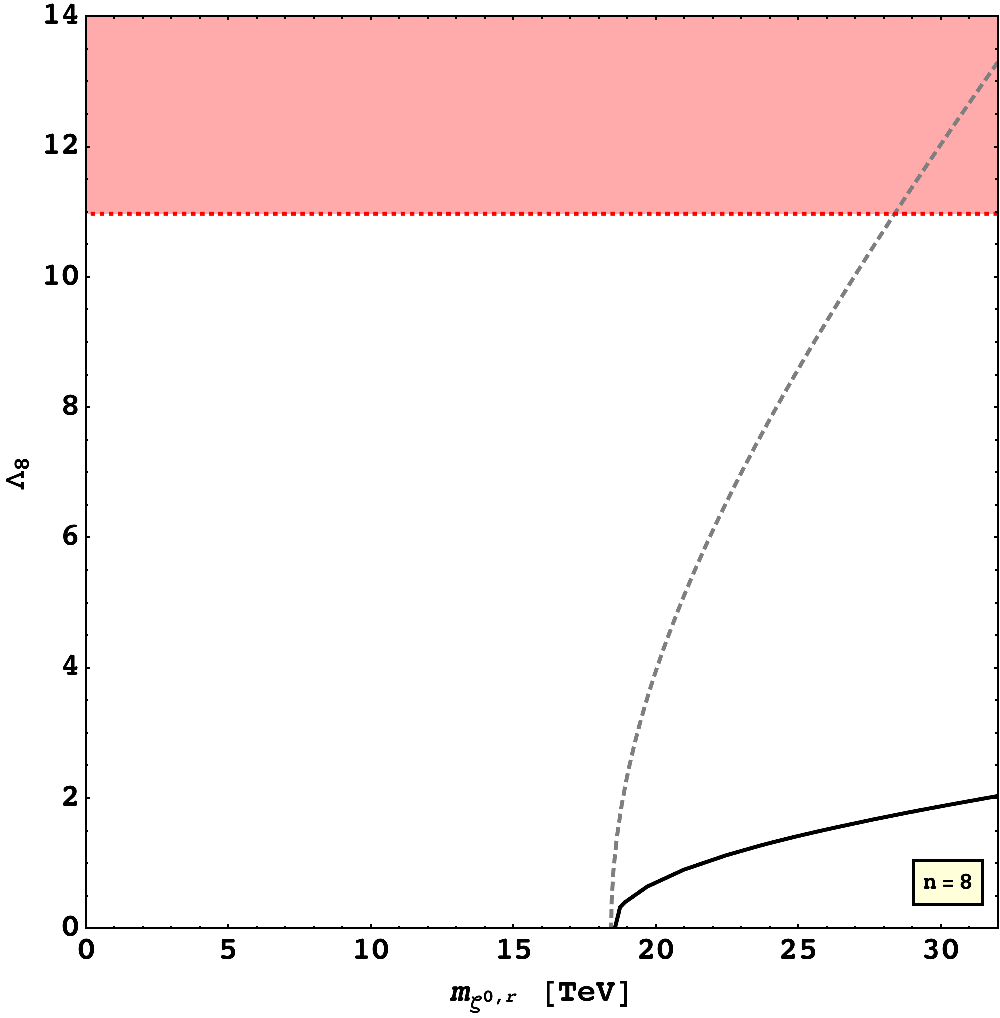}
\caption{Parameter values for which $\nOZR = 1$ for Sommerfeld enhancement from a Higgs-exchange Hulth\'{e}n potential (black solid) and single-species annihilation without the Sommerfeld effect (dashed grey), as a function of both $\mr$ and $\Lambda_n$. The region above the horizontal dotted red line is excluded by the unitarity bound, $\Lambda_n \leq \Lambda_n^{\rm MAX}$.}\label{fig:DM_frac_somm}
\end{center}
\end{figure}
\end{center}

We now calculate the effect of Sommerfeld enhancement on the relic abundance of $\zr$. We have seen that we can factor out the cross section from the thermal averaging, and so the DM fraction,~\cref{eq:DM_frac}, will be
\begin{equation}\label{eq:DM_frac_somm}
\OZR = \frac{\svSTD}{\left<\sigma\beta(\zr\,\zr\,\to\,W^{+}W^{-},\,ZZ,\,hh,\,f\bar{f})\right>\left<\Somm_h\right>}\;,
\end{equation}
where, again, we use $x = 20$ and $\svSTD = 3 \times 10^{-26}\,\unit{\frac{cm^3}{s}}$. We plot the result in~\cref{fig:DM_frac_somm}. The dashed grey curve shows where $\nOZR = 1$ using~\cref{eq:DM_frac_naive}, while the solid black line corresponds to $\nOZR = 1$ for the Sommerfeld-enhanced case,~\cref{eq:DM_frac_somm}. The region above the horizontal dotted red line (where $\Lambda_n^{\rm MAX} = \frac{n}{2}|\lambda_4^{\rm MAX}|$) is ruled out by the unitarity bound from~\cref{tbl:unitarity_lambdas}.

We saw in~\cref{fig:dm_multi_TeV_therm_somm} that the Sommerfeld enhancement factor increases with the $\zr\zr h$ coupling, $\Lambda_n$, so we are not surprised that its $\nOZR = 1$ curve coincides with the one from~\cref{eq:DM_frac_naive} at $\Lambda_n = 0$. For large values of $\Lambda_n$, the thermally-averaged Sommerfeld enhancement factor can be quite large. This will drive the cross section up, and thus the DM fraction down. Because of this, the range of allowed masses corresponding to $\nOZR = 1$ is $10.1$~TeV~$\leq \mr \lesssim 325$~TeV in $n=6$ and $18.5$~TeV~$\leq \mr \lesssim 400$~TeV in $n=8$. We would now like to combine the effects of co-annihilation and Sommerfeld enhancement.

\subsection{Combining co-annihilation with Sommerfeld enhancement}\label{sec:coann_somm}

The simple picture of a multiplicative enhancement factor or a sum of cross sections can no longer be applied when we consider Sommerfeld enhancement with the full large multiplet. For example, if we start with $\zp{+3}\HkQ{1}{-2}$ in the $n=6$ model, then the two initial-state particles might exchange a $W$ boson, changing to $\HkQ{1}{+2}\HkQ{1}{-1}$. They could then exchange a $Z$ boson, becoming $\HkQ{2}{+2}\HkQ{1}{-1}$. Exchanges of this sort continue until the final particles annihilate into, say, $W^{+}h$. The factorizing of the Sommerfeld effect in the single-particle case cannot be applied to this situation. For the Sommerfeld-enhanced co-annihilation $s_is_j \to s_k s_\ell \to$~SM$_A$ SM$_B$, $s_k s_\ell$ must run over all possible intermediate states.

In this case, we would normally be required to promote Schr\"{o}dinger's equation (\cref{eq:simple_schrodinger}) to a matrix equation. The potential will then take into account all possible exchanges as well as the mass splittings among the states. Generally, this does not have a closed-form solution and must be solved numerically. Due to the large number of states in our multiplet, this calculation quickly becomes intractable. However, as in the Inert Doublet case studied in Ref.~\cite{Garcia-Cely:2015khw}, if the DM follows the standard thermal freeze-out, in the high-DM-mass region the freeze-out will occur before the electroweak phase transition. In this regime the mass splittings vanish and we can work in a basis in which the couplings take a very simple form, dependent only on the total isospin and hypercharge of the two-particle initial state.

The electroweak phase transition (EWPT) occurs at $T_{\rm EWPT} \sim 0.2$~TeV~\cite{Laine:1998jb}. The mass range of our multi-TeV parameter space, $m \sim (5,\,50)$~TeV corresponds to a freeze-out termperature of $T_{\rm f.o.} = \nicefrac{m}{x} \sim (\nicefrac{5}{20},\,\nicefrac{50}{20})$~TeV~$= (0.25,\,2.5)$~TeV. Since $T_{\rm f.o.} > T_{\rm EWPT}$ over this whole range, the freeze-out of $\zr$ occurs before the EWPT---before electroweak symmetry is broken. This means that weak isospin and hypercharge are conserved quantities, they are ``good" quantum numbers. In addition, since the Higgs doublet has not yet acquired its vacuum expectation value, the states of $\mult$ are degenerate in mass. Furthermore, $W$ and $Z$ are massless, and the $h\mult\mult$ couplings ($\propto v$) all go to zero: we can use the Coulomb potential rather than the Yukawa or the Hulth\'{e}n potential. In this case, the potential matrix has elements $V_{\tau\varphi}$, where $\tau$ is the total isospin of the two-particle initial state and $\varphi$ is the total hypercharge. It is given by
\begin{equation}
V_{\tau\varphi} = \frac{\alpha_{\tau\varphi}}{r}\;.
\end{equation}
Whether this potential is repulsive or attractive will depend on the sign of $\alpha_{\tau\varphi}$, as defined below. The coupling parameter $\alpha_{\tau\varphi}$ may be determined for a generic $SU(2)_L \otimes U(1)_Y$ multiplet from Ref.~\cite{Strumia:2008cf}, where we find
\begin{equation}
\alpha_{\tau\varphi} = \left[2\tau(\tau+1) + 1 - n^2\right]\frac{g^2}{16\pi} + \eta_{\tau,\varphi}\,\frac{g^{\prime2}}{16\pi}\;,
\end{equation}
where $\eta_{0,0} = \eta_{1,0} = -1$, $\eta_{1,2} = +1$, and $\eta_{2,0} = 0$. Note that $\alpha_{\tau\varphi}$, and hence $\Somm_{\tau\varphi}$, is independent of $\lambda_{2,3,4}$. Since the exchanged bosons are all massless ($v = 0$), we can use the Sommerfeld enhancement factor for the Coulomb potential,
\begin{equation}\label{eq:somm_iso_combos}
\Somm_{\tau\varphi} = -\frac{\pi\,\alpha_{\tau\varphi}}{\beta}\,\frac{1}{1 - \e{\frac{\pi\,\alpha_{\tau\varphi}}{\beta}}}\;.
\end{equation}

The possible $(\tau,\varphi)$ combinations are shown in~\cref{tbl:dm_multi_TeV_coann_somm_iso}. The annihilation amplitudes are given by\footnote{In Ref.~\cite{Hally:2012pu}, $a_0([\mult\co \mult]_2 \to [WW]_2)$ is referred to as $a_0^{\perp}$.}~\cite{Earl:2013fpa,Hally:2012pu}
\begin{equation}\begin{aligned}\label{eq:a0_t0}
a_0([\mult\co\mult]_0 \to [\Phi\co\Phi]_0) &= - \frac{\sqrt{n}}{8 \sqrt{2} \pi}\lambda_2\;,\\
a_0([\mult\co\mult]_0 \to [WW]_0) &= \frac{g^2}{16 \pi} \frac{(n^2-1) \sqrt{n}}{2 \sqrt{3}}\;,\\
a_0([\mult\co\mult]_0 \to [BB]_0) &= \frac{g^2}{16 \pi} \frac{\ssw}{\ccw} \frac{Y_\mult^2 \sqrt{n}}{2}\;,
\end{aligned}\end{equation}
\begin{equation}\begin{aligned}\label{eq:a0_t1}
a_0([\mult\co\mult]_1 \to [\Phi\co\Phi]_1) &= - \frac{\sqrt{n(n^2-1)}}{32 \sqrt{6} \pi} \lambda_3\;,\\
a_0([\mult\co\mult]_1 \to [WB]_1) &= \frac{g^2}{16 \pi} \frac{\sw}{\cw} \frac{Y_\mult \sqrt{n (n^2 - 1)}}{\sqrt{6}}\;,\\
a_0([\mult\mult]_1 \to [\Phi \Phi]_1) &= -\frac{\sqrt{n(n^2-1)}}{16\sqrt{6} \pi} \lambda_4\;,
\end{aligned}\end{equation}
\begin{equation}\begin{aligned}\label{eq:a0_t2}
a_0([\mult\co \mult]_2 \to [WW]_2) &= \frac{g^2}{16\pi}\sqrt{\frac{n(n^2-1)(n^2-4)}{30}}\;.
\end{aligned}\end{equation}
Here $s_W$ and $c_W$ are the sine and cosine of the weak mixing angle and $g$ is the SU(2)$_L$ gauge coupling.
The elements of the annihilation cross section matrix, $\Gamma_{\tau\varphi}$, are then
\begin{equation}\begin{aligned}\label{eq:dm_multi_TeV_coann_somm_Gamma_IY}
\Gamma_{00} &= \frac{8\pi}{\mr^2}\Big[\Big|a_0([\mult\co\mult]_0 \to [\Phi\co\Phi]_0)\Big|^2 + \Big|a_0([\mult\co\mult]_0 \to [WW]_0)\Big|^2 + \Big|a_0([\mult\co\mult]_0 \to [BB]_0)\Big|^2\Big]\;,\\
\Gamma_{10} &= \frac{8\pi}{\mr^2}\Big[\Big|a_0([\mult\co\mult]_1 \to [\Phi\co\Phi]_1)\Big|^2 + \Big|a_0([\mult\co\mult]_0 \to [WB]_1)\Big|^2\Big]\;,\\
\Gamma_{12} &= \frac{8\pi}{\mr^2}\,\Big|a_0([\mult \mult]_1 \to [\Phi\Phi]_1)\Big|^2\;,\\
\Gamma_{20} &= \frac{8\pi}{\mr^2}\,\Big|a_0([\mult\co \mult]_2 \to [WW]_2)\Big|^2\;.
\end{aligned}\end{equation}

\begin{center}
\begin{table}[t]
\begin{center}
{\def\arraystretch{1.33}
\begin{tabular}{C{2.5em} C{2.5em} | C{16em}}
\hline\hline
$\tau$ & $\varphi$ & States\\
\hline
0 & 0 & $[\Phi\co\Phi]_{0}$, $[WW]_{0}$, $[BB]_{0}$, $[\mult\co\mult]_0$\\
1 & 0 & $[\Phi\co\Phi]_{1}$, $[WB]_{1}$, $[\mult\co\mult]_1$\\
1 & 2 & $[\Phi\Phi]_{1}$, $[\mult \mult]_1$\\
2 & 0 & $[WW]_{2}$, $[\mult\co \mult]_2$\\
\hline\hline
\end{tabular}}
\caption{Properly-normalized total isospin and hypercharge combinations in the unbroken SM. These combinations are used in the calculation of the co-annihilating Sommerfeld-enhanced relic abundance. Explicit expressions for each combination are given in~\cref{sec:iso_combos}.}\label{tbl:dm_multi_TeV_coann_somm_iso}
\end{center}
\end{table}
\end{center}

\begin{center}
\begin{figure}[t]
\begin{center}
\includegraphics[width=0.45\textwidth]{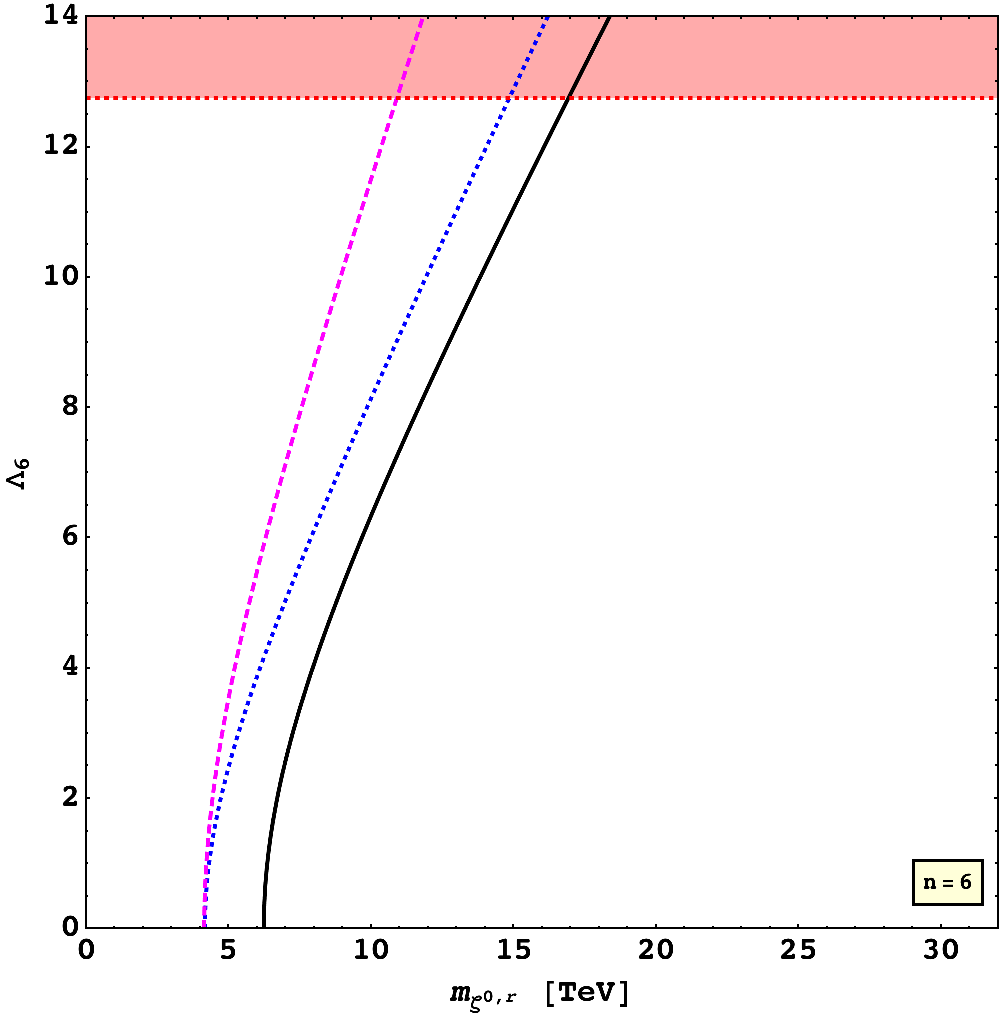}\hspace*{0.67em}
\includegraphics[width=0.45\textwidth]{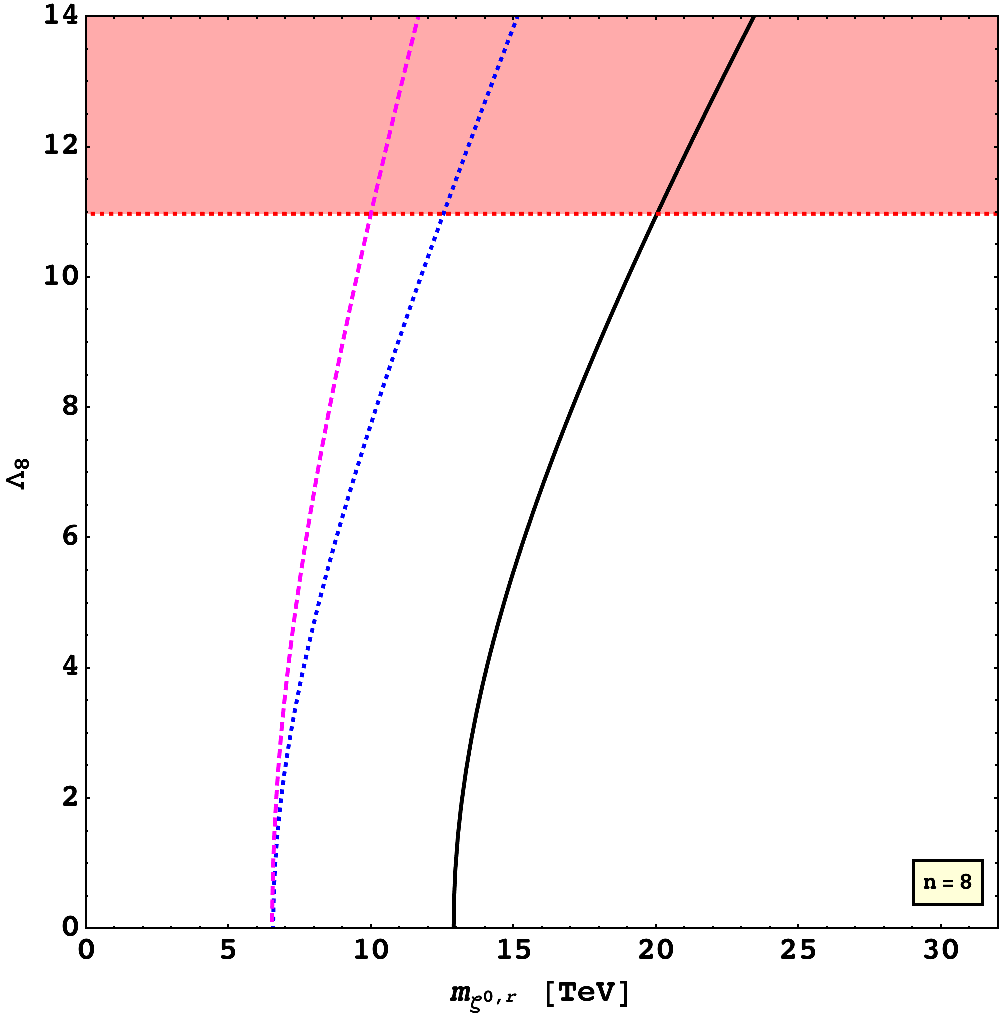}
\caption{Parameter values for which $\nOZR = 1$ comparing co-annihliation and Sommerfeld enhancement. The left-most (dashed magenta) curve corresponds to setting $\Somm_{\tau\varphi} = 1$ in~\cref{eq:somm_co_iso_combos} and allows us to see the effect of setting $v = 0$. The middle (dotted blue) curve corresponds to $\nOZR = 1$ for co-annihilations as in~\cref{fig:DM_frac_coann}. The right-most (solid black) curve corresponds to the full expression in~\cref{eq:DM_frac_somm_coann}.}\label{fig:DM_frac_coann_comp}
\end{center}
\end{figure}
\end{center}

Combining the Sommerfeld enhancement,~\cref{eq:somm_iso_combos}, with the annihilation cross sections,~\cref{eq:dm_multi_TeV_coann_somm_Gamma_IY}, the total Sommerfeld-enhanced co-annihilating cross section is given by~\cite{Garcia-Cely:2015khw}
\begin{equation}\label{eq:somm_co_iso_combos}
(\sigma\beta)_{\Somm} = \frac{2}{(2n)^2}\,\sum_{\tau,\varphi}\left(2\tau + 1\right)\,\Gamma_{\tau\varphi}\,\Somm_{\tau\varphi}\;,
\end{equation}
where the factor of 2 in the numerator accounts for the normalization of the amplitudes in Eqs.~(\ref{eq:a0_t0}--\ref{eq:a0_t2}) and the factor $(2\tau + 1)$ is the multiplicity of each isospin state. The DM fraction is then given by
\begin{equation}\label{eq:DM_frac_somm_coann}
\OZR = \frac{\svSTD}{\left<(\sigma\beta)_{\Somm}\right>}\;.
\end{equation}

To determine the DM fraction, we set $\lambda_2 = \lambda_3 = 0$ and $\lambda_4 = \frac{2}{n}(-1)^{\frac{n}{2}}\Lambda_n$, scan over $\mr$ and $\Lambda_n$, thermally average $\Somm_{\tau\varphi}$ at each point with $x \equiv m_{\zeta^{0,r}}/T = 20$, and compare to $\svSTD$ using~\cref{eq:DM_frac_somm_coann}. We examine the various effects in~\cref{fig:DM_frac_coann_comp}. Using~\cref{eq:DM_frac_coann}, we get the dotted blue curve, reproducing~\cref{fig:DM_frac_coann}. We then take~\cref{eq:somm_co_iso_combos} and set $\Somm_{\tau\varphi} = 1$, resulting in the dashed magenta curve. This shows the effect of setting $v = 0$ in the co-annihilation calculation,~\cref{eq:DM_frac_coann}. Finally, we plot the full~\cref{eq:DM_frac_somm_coann}, which gives the solid black curve in~\cref{fig:DM_frac_coann_comp}.

\begin{center}
\begin{figure}[t]
\begin{center}
\includegraphics[width=0.45\textwidth]{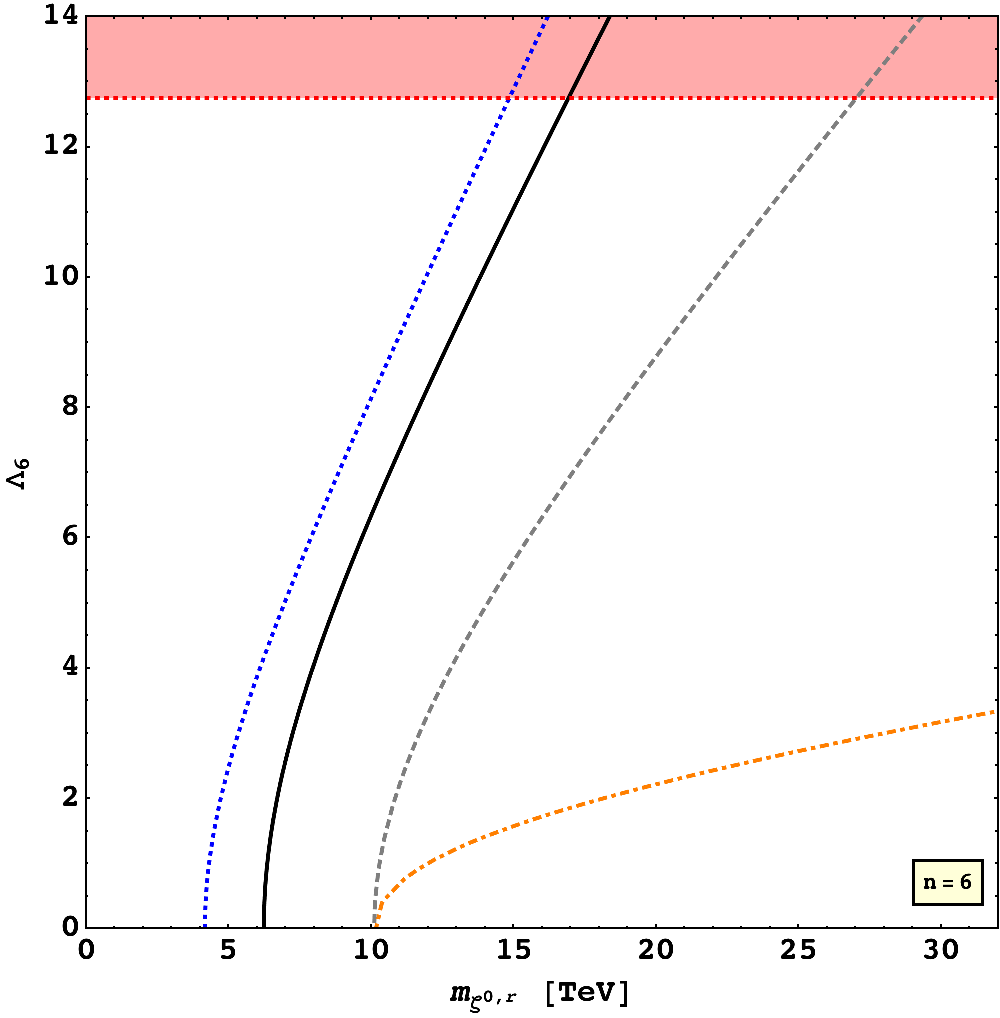}\hspace*{0.67em}
\includegraphics[width=0.45\textwidth]{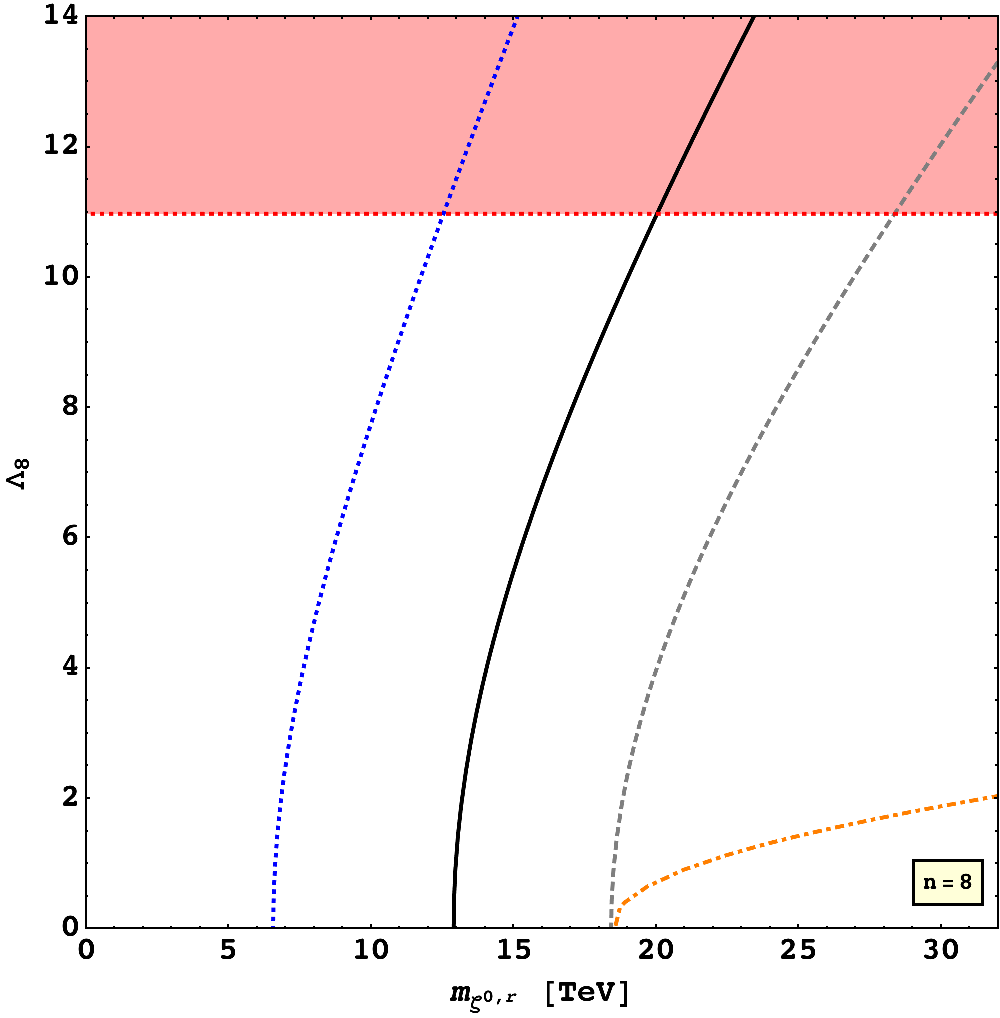}
\caption{Parameter values for which $\nOZR = 1$, with $n=6$ on the left and $n=8$ on the right. The solid black line incorporates both co-annihilations and Sommerfeld enhancement (our final results,~\cref{eq:DM_frac_somm_coann}). The dot-dashed orange line shows Sommerfeld enhancement with only a Higgs potential (\cref{eq:DM_frac_somm}). The dotted blue line shows co-annihilations (\cref{eq:DM_frac_coann}). The dashed grey line shows the case of single-species annihilation without Sommerfeld effects (\cref{eq:DM_frac_naive}). The region above the horizontal dotted red line is excluded by the unitarity bound, $\Lambda_n \leq \Lambda_n^{\rm MAX}$.}\label{fig:DM_frac_all}
\end{center}
\end{figure}
\end{center}

To summarize, in~\cref{fig:DM_frac_all}, we plot the $\nOZR = 1$ curves for the four cases of interest. The dashed grey curve is the na\"{i}ve DM fraction (no co-annihilation and no Sommerfeld enhancement), the dotted blue line is the co-annihilating DM fraction (no Sommerfeld enhancement), the dot-dashed orange curve is the Sommerfeld-enhanced DM fraction (no co-annihilation), and the solid black curve is the Sommerfeld-enhanced co-annihilating cross section (the full calculation). Due to the factor of $\nicefrac{1}{(2n)^2}$ in the co-annihilation case, the DM fraction of the Sommerfeld-enhanced co-annihilating case is increased, pushing the allowed mass range to lower values. The lack of Sommerfeld enhancement from Higgs exchange (since we worked in $T_{\rm f.o.} > T_{\rm EWPT}$ where $v = 0$) means that the Sommerfeld enhancement factor itself in this situation does not depend on $\Lambda_n$, and so is constant for given values of $\tau$ and $\varphi$.  This is why the Sommerfeld-enhanced co-annihilating curves (solid black in Fig.~\ref{fig:DM_frac_all}) are roughly the same shape as the co-annihilation curves without Sommerfeld enhancement (dotted blue). We find values $\left<\Somm\right> \sim 2 - 2.7$ for the combinations of total weak isospin and hypercharge that are required here.  The solid curves in Fig.~\ref{fig:DM_frac_all} are our final results.

In summary, then, as $\mr$ gets large, we reach a point where $\nOZR = 1$. In this region, the parameter space is constrained only by perturbative unitarity of $\lambda_{2,3,4}$. However, other effects such as co-annihilation and Sommerfeld enhancement will alter the DM fraction of $\zr$, leading to different allowed mass ranges. When combining Sommerfeld enhancement and co-annihilations, we find that the allowed mass range is $6.2$~TeV~$\leq \mr \leq 16.9$~TeV in the $n=6$ case and $12.9$~TeV~$\leq \mr \leq 20.0$~TeV in the $n=8$ case.  These masses are almost a factor of 2 lower than the na\"ive predictions not including co-annihilation and Sommerfeld effects, and constitute an upper bound on the mass of the large multiplet to avoid over-closing the Universe (assuming a standard thermal history).

\section{Landau poles in the high-mass region of the large multiplet models}\label{sec:landau_pole}

The quartic couplings of scalar field theories typically increase with increasing mass scale due to renormalization group running, leading to an eventual divergence called a Landau pole.  This indicates a breakdown of the theory, requiring new physics at or below the scale of the Landau pole.  Theories that contain scalars in large gauge-group representations are known to run faster due to the large multiplicity of states (see, e.g., Ref.~\cite{Hamada:2015bra}).  The large scalar multiplet model with $n=6$ and $Y_{\Sigma} = 1$ was studied in Ref.~\cite{Hamada:2015bra}; we reproduce their results and extend them to the $n = 8$ model\footnote{The real scalar multiplet with $n = 7$ was also studied in Ref.~\cite{Cai:2015kpa}, which found that the addition of Yukawa interactions between the scalar multiplet and exotic fermions can push the scale of the Landau pole much higher than in the simple scalar extensions considered here.}.  The one-loop renormalization group equations (RGEs) in our scalar potential parameterization are given in~\cref{sec:RGEs}.

To compute the scale of the Landau pole for the scenarios in which the $n=6$ or $8$ model accounts for all the dark matter, we set the initial conditions for the RGE running at the low scale $\mu_0$ to be\footnote{For the numerical calculation we take $m_h = 125$~GeV, $v = 246$~GeV, $\alphaem = \nicefrac{1}{128}$, $\ssw = 0.231$, $\alphas = 0.1185$, and $m_t = 173$~GeV.}
\begin{equation}\begin{aligned}\label{eq:rge_init_cond_nonzero}
\mu_0 &= \mr\;,&&&
\lambda_1(\mu_0) &= \frac{m_h^2}{2v^2}\;,\\
\lambda_{i \neq 1,4}(\mu_0) &= 0\;,&&&
\lambda_4(\mu_0) &= \lambda_4\;,\\
g_1(\mu_0) &= \sqrt{\frac{5}{3}}\,\frac{\sqrt{4\pi\alphaem}}{\cw}\;,&&&
g_2(\mu_0) &= \frac{\sqrt{4\pi\alphaem}}{\sw}\;,\\
g_3(\mu_0) &= \sqrt{4\pi\alphas}\;,&&&
y_t(\mu_0) &= \sqrt{2}\frac{m_t}{v}\;,\\
m^2(\mu_0) &= -\frac{m_h^2}{2}\;,&&&
M^2(\mu_0) &= \mr^2 + \frac{1}{2}v^2\,\Lambda_n\;,
\end{aligned}\end{equation}
where $g_1$ is the hypercharge coupling in the grand-unified theory (GUT)-normalization, $g_2$ and $g_3$ are the SU(2)$_L$ and strong-interaction couplings, $\alpha_{\rm EM}$ is the electromagnetic fine structure constant, $\alpha_{\rm s} = g_3^2/4 \pi$ is the strong-interaction equivalent, and $y_t$ and $m_t$ are the top quark Yukawa coupling and mass, respectively.

\begin{center}
\begin{figure}[t]
\begin{center}
\includegraphics[width=0.45\textwidth]{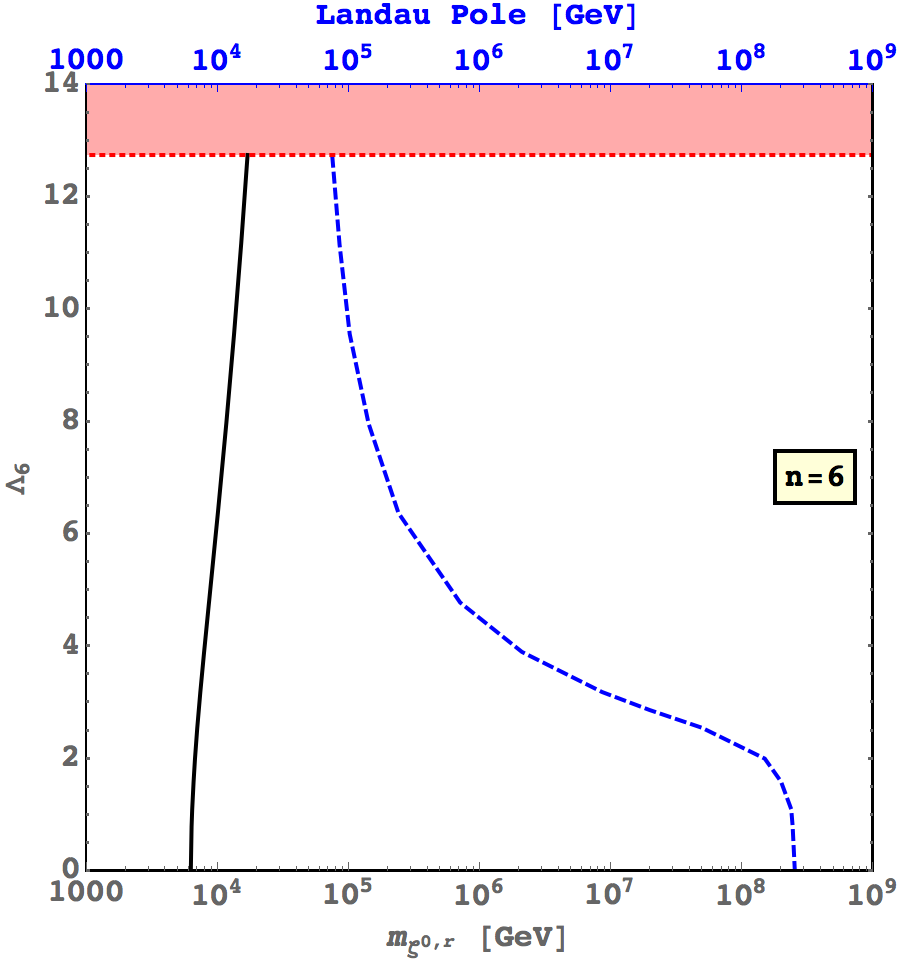}\hspace*{0.67em}
\includegraphics[width=0.45\textwidth]{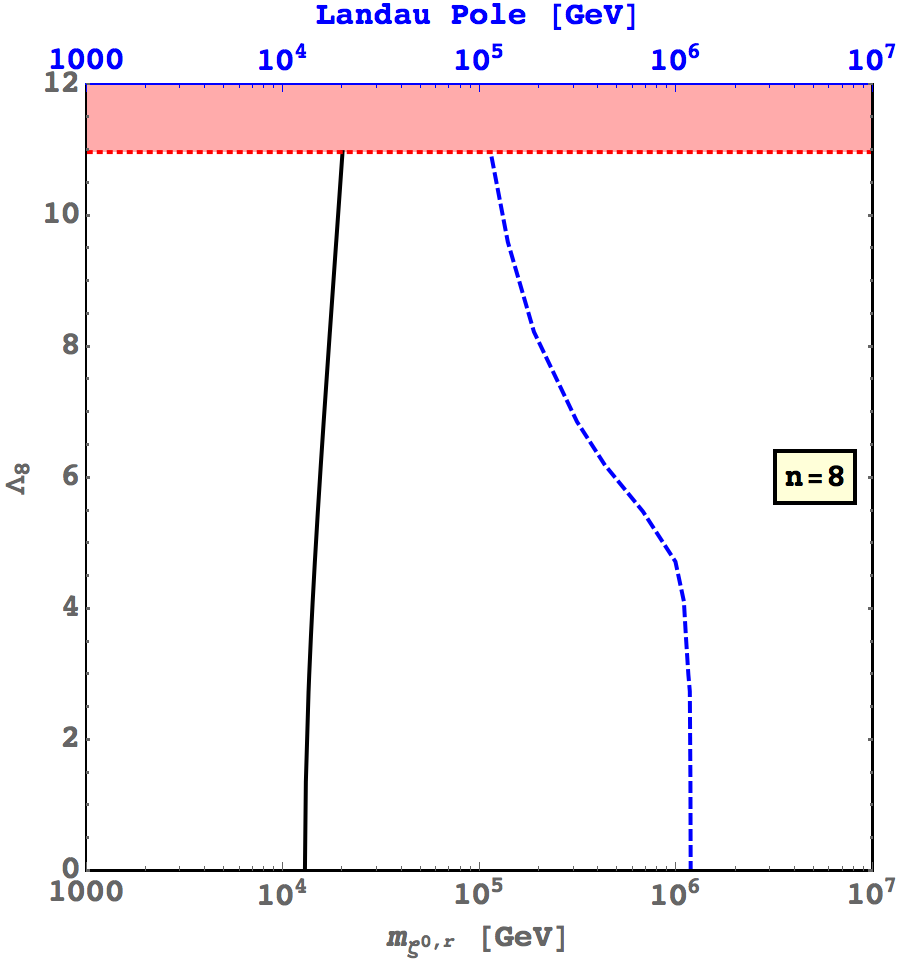}
\caption{Location of the (one-loop) Landau pole in the $n = 6$ (left panel) and $n = 8$ (right panel) large multiplet models. The solid black curve shows $\mr$ (bottom axis) for which $\nOZR = 1$ as in~\cref{fig:DM_frac_all}, while the dashed blue curve shows the location of the Landau pole (top axis) for the corresponding parameter point.  We set $\lambda_2 = \lambda_3 = 0$ at the $\mr$ scale as usual.}\label{fig:landau_pole_locs}
\end{center}
\end{figure}
\end{center}

We vary $\mr$ as a function of $\lambda_4$ to yield the correct DM relic abundance as found in~\cref{sec:coann_somm}. 
We then numerically solve the RGEs to determine where $\lambda^{-1} = 0$, the location of the Landau pole. In~\cref{fig:landau_pole_locs}, we plot both the Sommerfeld enhanced co-annihilation DM fraction (solid black curve) as well as the location of the Landau pole (dashed blue). We see that for relatively small quartic coupling ($|\lambda_4|~\lesssim 1$), the Landau pole remains roughly 4 (2) orders of magnitude above the scalar masses in the $n = 6$ ($8$) model. As $|\lambda_4|$ increases beyond this value, it begins to contribute significantly to the initial RGE running and causes the Landau pole to occur at much lower energy; for $\lambda_4$ at the unitarity bound, the Landau pole occurs less than an order of magnitude above $\mr$ in either model. This indicates that, if our models are coupled such that $\lambda_4$ is near the unitarity bound, some other form of new physics must necessarily also be present.

\begin{center}
\begin{figure}[t]
\begin{center}
\includegraphics[width=0.45\textwidth]{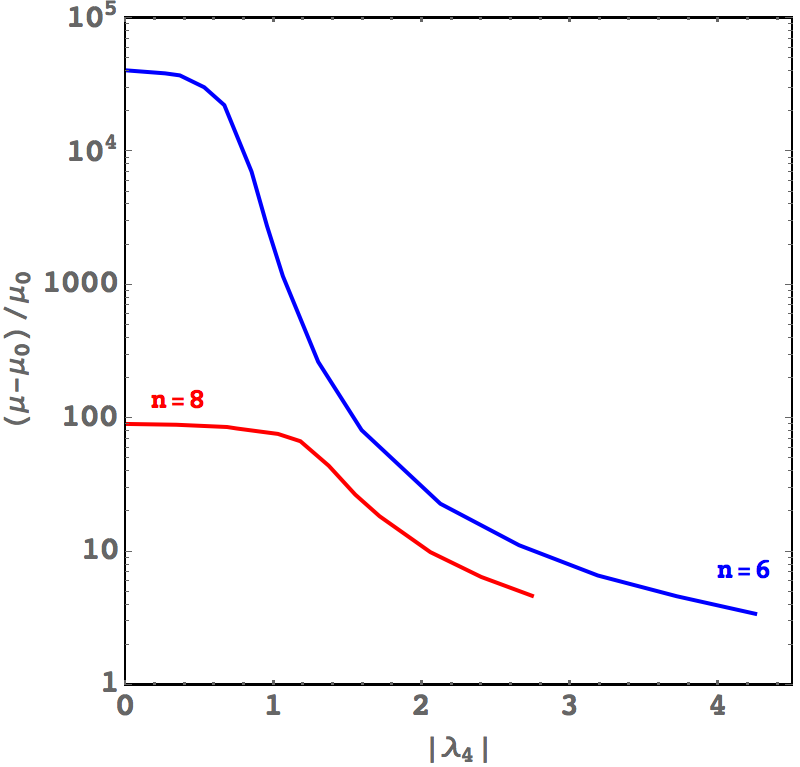}
\caption{Normalized difference between the Landau pole scale $\mu$ and the $\zr$ mass scale $\mu_0$ as a function of $|\lambda_4|$. In these figures, we set the initial conditions as in~\cref{eq:rge_init_cond_nonzero}. The upper blue curve corresponds to $n=6$ and the lower red curve to $n=8$. The endpoint of each curve occurs at the unitarity bound on $|\lambda_4|$ from~\cref{tbl:unitarity_lambdas}.}\label{fig:landau_pole_ratio}
\end{center}
\end{figure}
\end{center}

To highlight the dependence of the Landau pole location on the initial conditions, we plot the difference between the Landau-pole scale, $\mu$, and initial scale, $\mu_0$, normalized by $\mu_0$ in~\cref{fig:landau_pole_ratio}. In this figure, we choose $\mr$ such that the correct relic abundance is obtained, as detailed in~\cref{sec:coann_somm}. The upper blue curve gives the ratio for $n=6$, while the lower red curve is for $n=8$. The endpoint of each curve corresponds to the unitarity bound given in~\cref{tbl:unitarity_lambdas}.

\section{Dark matter direct detection prospects}\label{sec:direct_detection}

The scattering of a $\zr$ off of a nucleus proceeds only via Higgs exchange. The resulting spin-independent per-nucleon cross section is~\cite{Earl:2013fpa}
\begin{equation}\label{eq:dm_dd_scatt_sigma}
\sigma_{{\rm SI},N}^{\zeta} = \frac{f_N^2\,\Lambda_n^2}{4\pi}\,\frac{v^2}{m_h^4}\,\frac{m_N^2}{(\mr + m_N)^2}\;,
\end{equation}
where~\cite{Ellis:2000ds,Cheng:2014opa}
\begin{equation}\begin{aligned}\label{tbl:dm_dd_H_n_p_f}
f_p &= \frac{m_p}{v}\left(0.350 \pm 0.048\right)\;,\\
f_n &= \frac{m_n}{v}\left(0.353 \pm 0.049\right)\;,
\end{aligned}\end{equation}
for protons and neutrons, respectively. Since $f_n \approx f_p$ within uncertainties, we will use $f_N = 0.35$ in~\cref{eq:dm_dd_scatt_sigma}.

\begin{center}
\begin{figure}[t]
\begin{center}
\includegraphics[width=0.5\textwidth]{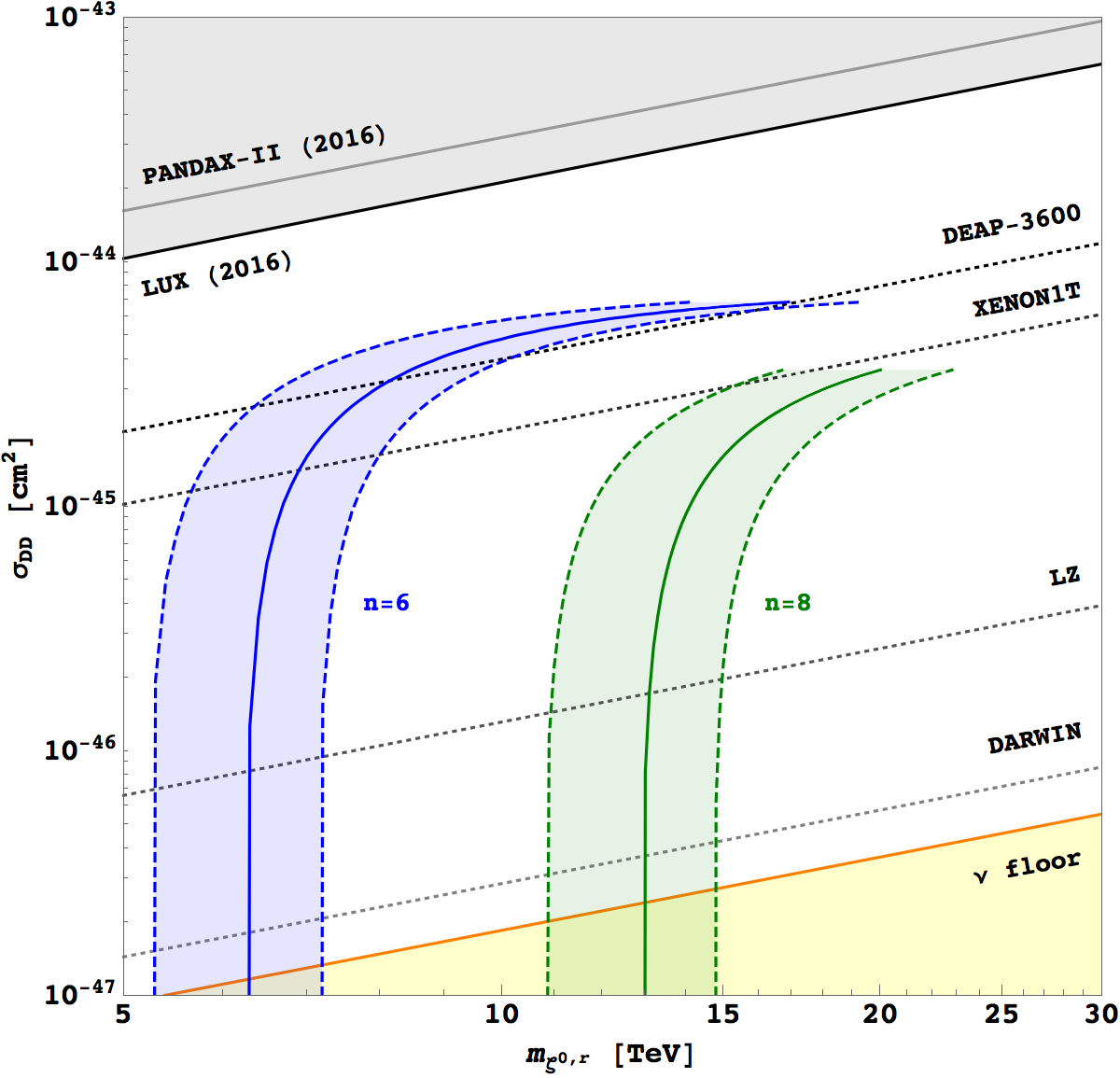}
\caption{Direct detection predictions for the multi-TeV parameter space in the $n=6$ (blue) and $n=8$ (green) models. The solid lines show where $\nOZR = 1$ in each model, incorporating both Sommerfeld enhancement and co-annihilations. The dashed lines on either side show the range of direct detection cross section for $\nOZR \in [0.7,\,1.3]$ to account for the uncertainty in $\svSTD$. The solid grey curve shows the extrapolation of the limit from the PandaX-II experiment~\cite{Tan:2016zwf}, while the solid black curve shows the limit from LUX~\cite{Akerib:2016vxi}. The dotted grey curves show extrapolations of the projected sensitivity of upcoming experiments, from top to bottom: DEAP-3600~\cite{Amaudruz:2014nsa}, XENON1T~\cite{Aprile:2015uzo}, LZ~\cite{McKinsey:2016xhn}, and DARWIN~\cite{Aalbers:2016jon}. The shaded orange region at the bottom of the plot shows where the coherent scattering of neutrinos will become a non-negligible background, from Ref.~\cite{Billard:2013qya}.}\label{fig:dm_high_m0_directdetection}
\end{center}
\end{figure}
\end{center}

Using the Sommerfeld-enhanced co-annihilating $\nOZR = 1$ curve from~\cref{fig:DM_frac_all}, we determine $\Lambda_n \equiv \Lambda_n(\mr)$ and use this to evaluate~\cref{eq:dm_dd_scatt_sigma}. We plot the resulting curve up to the unitarity bound in~\cref{fig:dm_high_m0_directdetection} for $n=6$ (blue curve) and $n=8$ (green curve). To account for the uncertainty in $\svSTD$, we also calculate $\sigma^{\rm expt.}_{\rm SI} \in [0.7\,\sigma_{\rm SI}^{\zeta},\,1.3\,\sigma_{\rm SI}^{\zeta}]$ as a function of $\mr$ and plot this as the shaded region between the pairs of dashed curves.

The exclusion limits provided by experiments typically only go up to $M_{\rm DM} \sim 1$~TeV (although the limit from LUX extends to $100$~TeV~\cite{Akerib:2016vxi} and the projection from DEAP-3600 goes up to $12.5$~TeV~\cite{Amaudruz:2014nsa}). Because of this, we extrapolate the others up to the multi-TeV region of interest using a linear fit for the expected $\sigma_{\rm DD} \propto M_{\rm DM}$ rise at high mass. In~\cref{fig:dm_high_m0_directdetection}, the solid black line shows the current best exclusion limit from LUX~\cite{Akerib:2016vxi} and the solid grey line corresponds to an extrapolation of the limit from the PandaX-II experiment~\cite{Tan:2016zwf}. The dotted grey lines correspond to extrapolations of the projected future limits provided by (from top to bottom): DEAP-3600~\cite{Amaudruz:2014nsa}, XENON1T~\cite{Aprile:2015uzo}, LZ~\cite{McKinsey:2016xhn}, and DARWIN~\cite{Aalbers:2016jon}. The shaded orange region at the bottom of the plot corresponds to where the coherent scattering of neutrinos produced in cosmic ray collisions in the atmosphere and neutrinos produced in core-collapse supernovae becomes an irreducible background (this is calculated up to $\sim10$~TeV in Ref.~\cite{Billard:2013qya} and we extrapolate it in the same manner as the other projections).

Both the $n=6$ and $n=8$ models evade the current experimental limits. The projected sensitivity of DEAP-3600 will allow us to probe the part of the $n = 6$ model near the unitarity bound. XENON1T will probe the $n=6$ model for $\Lambda_6 \gtrsim 4$ and will begin to probe the $n=8$ model near the unitarity bound. The LZ experiment will probe a large fraction of the remaining parameter space in both models. The proposed DARWIN experiment would extend this even further. As $\Lambda_n$ becomes small, the cross section drops rapidly, and the neutrino background will become important for $\Lambda_n \lesssim 0.7$.

\section{Discussion and conclusions}\label{sec:summary}

In this paper we examined the $Z_2$-symmetric large scalar multiplet models with $T$ = 5/2 and 7/2 in the multi-TeV $\mr$ region of parameter space, in which the lightest $Z_2$-odd scalar can constitute all of the dark matter.  This completes the study of \emph{all} perturbative DM models that extend the SM Higgs sector by a single (inert) scalar multiplet.  In this high-mass region, the only pre-existing constraints on the parameter space are those arising from the absence of alternate minima ($M^2 > 0$) and perturbative unitarity of the quartic scalar couplings (\cref{tbl:unitarity_lambdas}).  In calculating the relic density in this high-mass range we must take into account co-annihilations and Sommerfeld enhancement, which together reduce the allowed masses for $\zr$ to constitute all the DM by almost a factor of two. We find that the allowed mass range for $\zr$ to constitute all the DM is $6.2$~TeV~$\leq \mr \leq 16.9$~TeV in the $n=6$ case and $12.9$~TeV~$\leq \mr \leq 20.0$~TeV in the $n=8$ case, where the range of masses corresponds to the perturbative range of the $\zr\zr h h$ coupling $\Lambda_{n}$.  These masses constitute an upper limit on the mass of $\zr$ to avoid over-closing the Universe; to accommodate masses above these bounds would require additional new physics leading to a non-standard thermal history of the Universe.

We also investigated the scale of the Landau pole when these models account for all the dark matter.  We find that when the coupling $\lambda_4(\mu_0) = \frac{2}{n}(-1)^{\frac{n}{2}}\Lambda_n$ is small, the Landau pole occurs roughly 4 (2) orders of magnitude above the scalar masses in the $n = 6$ ($8$) model.  In particular, the Landau pole must occur below $3 \times 10^8$~GeV in the $n=6$ model and below about $10^6$~GeV in the $n=8$ model, indicating that these models must be ultraviolet-completed well below the Planck scale.  Higher $\lambda_4$ values bring down the Landau pole until it is only an order of magnitude above the scalar masses for $\lambda_4$ at its perturbative unitarity bound.

Dark matter direct-detection experiments such as DEAP-3600 and XENON1T, which have just begun their physics data-taking runs, will be able to probe the more strongly-coupled region of parameter space in the $n=6$ model. To explore the remainder of the parameter space (down to the neutrino floor) through direct detection will require next-generation multi-tonne experiments such as LZ and, ultimately, an experiment such as DARWIN.

An additional promising avenue to constrain or discover high-mass dark matter is through indirect detection of its annihilation products, including gamma rays, antiprotons, and positrons.  Particularly promising for multi-TeV dark matter are the Cherenkov gamma-ray detectors, including H.E.S.S.~\cite{Abramowski:2013ax}, HAWC~\cite{Abeysekara:2014ffg}, and CTA~\cite{Silverwood:2014yza}.  The sensitivity of indirect detection in the multi-TeV range is largely due to the Sommerfeld enhancement of dark matter annihilation at the relatively low collision velocities in galactic halos, leading to large resonant enhancements of the annihilation cross section for certain DM masses (as in the right panel of~\cref{fig:dm_multi_TeV_somm_enhance_hulth}).

The proper treatment of the Sommerfeld enhancement in today's galactic halos poses a significant computational challenge.  Due to the small mass splittings in multi-TeV DM, the Sommerfeld calculation involves $\zr \zr \to s_k s_\ell \to {\rm SM}_A {\rm SM}_B$ with all possible intermediate two-particle states $s_k s_\ell$ taken into account.  Unlike during freeze-out, we are below the electroweak phase transition and cannot simplify the calculation by taking the Higgs vev and all gauge boson masses to be zero, which allowed us to use closed-form solutions for the Sommerfeld enhancement factor.  Instead, the calculation of the Sommerfeld enhancement factor for each parameter point involves numerically solving a coupled set of Schr\"odinger equations for all possible two-particle intermediate states, which is beyond the scope of this paper.  We therefore leave this avenue to future work.

\begin{acknowledgments}

We thank Mary-Jean Harris for helping us understand the Sommerfeld effect.
This work was supported by the Natural Sciences and Engineering Research Council of Canada. 

\end{acknowledgments}


\appendix

\section{Generators and conjugation matrices}\label{sec:matrices}

For a complex scalar multiplet $\mult$ with hypercharge $Y_{\mult}=1$ (normalized so that $Q = T^3 + Y/2$), the most general gauge-invariant and $Z_2$-invariant renormalizable scalar potential was given in~\cref{eq:conjugatepotential}, in which $\widetilde \Phi =  i \sigma^2 \Phi^*$ and $\widetilde \mult = C \mult^*$ are the conjugate multiplets.  Here $\sigma^2$ is the second Pauli matrix and the conjugation matrix $C$ for the large multiplet is an anti-diagonal $n \times n$ matrix. For $n=6$ and 8, the matrix $C$ is given by
\begin{equation}
C_{(n = 6)} = \left(\begin{matrix} 0 & 0 & 0 & 0 & 0 & 1\\
0 & 0 & 0 & 0 & -1 & 0\\
0 & 0 & 0 & 1 & 0 & 0\\
0 & 0 & -1 & 0 & 0 & 0\\
0 & 1 & 0 & 0 & 0 & 0\\
-1 & 0 & 0 & 0 & 0 & 0
\end{matrix}\right),  \qquad
C_{(n = 8)} = \left(\begin{matrix}0 & 0 & 0 & 0 & 0 & 0 & 0 & 1\\
0 & 0 & 0 & 0 & 0 & 0 & -1 & 0\\
0 & 0 & 0 & 0 & 0 & 1 & 0 & 0\\
0 & 0 & 0 & 0 & -1 & 0 & 0 & 0\\
0 & 0 & 0 & 1 & 0 & 0 & 0 & 0\\
0 & 0 & -1 & 0 & 0 & 0 & 0 & 0\\
0 & 1 & 0 & 0 & 0 & 0 & 0 & 0\\
-1 & 0 & 0 & 0 & 0 & 0 & 0 & 0
\end{matrix}\right).
\label{appeq:Cmatrix}
\end{equation}

Taking $\lambda_4$ real and working in unitarity gauge, the term involving $\lambda_4$ in the scalar potential of Eq.~(\ref{eq:conjugatepotential}) reduces to
\begin{equation}
	\lambda_4\, \widetilde{\Phi}^\dag \tau^a \Phi\, \mult^\dag T^a \widetilde{\mult} + \mathrm{h.c.} 
	= \frac{1}{4}\lambda_4 (h + v)^2\left[\mult^\dag T^{-} \widetilde{\mult} 
	+ \widetilde{\mult}^\dag T^{+} \mult\right],
\end{equation}
where $T^{\pm} = T^1 \pm i T^2$. For $n=6$ the generators $T^a$ are given by
\begin{equation}
T_{(n = 6)}^{+} = \left(\begin{matrix}0 & \sqrt{5} & 0 & 0 & 0 & 0 \\
0 & 0 & 2 \sqrt{2} & 0 & 0 & 0 \\
0 & 0 & 0 & 3 & 0 & 0 \\
0 & 0 & 0 & 0 & 2 \sqrt{2} & 0 \\
0 & 0 & 0 & 0 & 0 & \sqrt{5} \\
0 & 0 & 0 & 0 & 0 & 0\end{matrix}\right) = \left(T_{(n = 6)}^{-}\right)^{\dagger},
\end{equation}
\begin{equation}
T_{(n = 6)}^3 = \frac{1}{2}\,\mathrm{diag}\left(5,\,3,\,1,\,-1,\,-3,\,-5\right),
\end{equation}
while for $n=8$ they are
\begin{equation}
T_{(n = 8)}^{+} = \left(\begin{matrix}0 & \sqrt{7} & 0 & 0 & 0 & 0 & 0 & 0 \\
0 & 0 & 2 \sqrt{3} & 0 & 0 & 0 & 0 & 0 \\
0 & 0 & 0 & \sqrt{15} & 0 & 0 & 0 & 0 \\
0 & 0 & 0 & 0 & 4 & 0 & 0 & 0 \\
0 & 0 & 0 & 0 & 0 & \sqrt{15} & 0 & 0 \\
0 & 0 & 0 & 0 & 0 & 0 & 2 \sqrt{3} & 0 \\
0 & 0 & 0 & 0 & 0 & 0 & 0 & \sqrt{7} \\
0 & 0 & 0 & 0 & 0 & 0 & 0 & 0\end{matrix}\right) = \left(T_{(n = 8)}^{-}\right)^{\dagger},
\end{equation}
\begin{equation}
T_{(n = 8)}^3 = \frac{1}{2}\,\mathrm{diag}\left(7,\,5,\,3,\,1,\,-1,\,-3,\,-5,\,-7\right).
\end{equation}

\section{Feynman rules}\label{sec:feynman_rules}

In this section we collect the Feynman rules for the couplings of the new scalars to gauge and Higgs bosons. We define the couplings with all particles and momenta incoming. For couplings involving scalar momenta, we define $p_1$ as the momentum of the first scalar and $p_2$ as the momentum of the second scalar. 

For simplicity in the derivation of the oblique parameters, all coefficients $C$ for couplings of scalars to one or two electroweak gauge bosons are defined with the overall factors of $e$ removed: one factor of $e$ is removed from couplings to a single gauge boson and two factors of $e$ are removed from couplings to two gauge bosons.

The full list of Feynman rules is give in Appendix B of Ref.~\cite{Earl:2013fpa}.

\subsection{Higgs boson couplings to scalar pairs}

The Feynman rule for the coupling of two new scalars to a Higgs boson, $h s_1 s_2$, is given by $-i C_{h s_1s_2}$, where
\begin{align}
C_{h \zr \zr} &= v \left(\lambda_2 + \frac{1}{4}\lambda_3 + \frac{n}{2} (-1)^{\frac{n}{2}+1} \lambda_4\right)\;,\nonumber\\
C_{h \zi \zi} &= v\left(\lambda_2 + \frac{1}{4}\lambda_3 + \frac{n}{2} (-1)^{\frac{n}{2}}\,\lambda_4\right)\;,\nonumber\\
C_{h \HkQ{1}{Q} \HkQ{1}{-Q}} &= v\left(\lambda_2 + \frac{1}{4}\lambda_3 - \frac{1}{2}\sqrt{Q^2\lambda_3^2 + (n^2-4Q^2)\lambda_4^2}\right)\;,\nonumber\\ 
C_{h \HkQ{2}{Q} \HkQ{2}{-Q}} &= v\left(\lambda_2 + \frac{1}{4}\lambda_3 + \frac{1}{2}\sqrt{Q^2\lambda_3^2 + (n^2-4Q^2)\lambda_4^2}\right)\;,\nonumber\\ 
C_{h \zp{\frac{n}{2}} \zp{-\frac{n}{2}}} &= v\left(\lambda_2 - \frac{2Q-1}{4}\lambda_3\right)\;.
\end{align}

\subsection{Gauge boson couplings to scalar pairs}

The Feynman rules for the couplings of the new scalars to gauge bosons come from the gauge-kinetic terms in the Lagrangian,
\begin{equation}
\mathcal{L} \supset \left( \mathcal{D}_\mu \mult \right)\ct \left( \mathcal{D}^\mu \mult \right),
\end{equation}
where the covariant derivative is given by
\begin{equation}
\mathcal{D}_\mu = \partial_{\mu} - i \frac{g}{\sqrt{2}} \left( W_{\mu}^+ T^+ + W_{\mu}^- T^- \right) - i \frac{e}{s_W c_W} Z_{\mu} \left( T^3 - s_W^2 Q \right) - i e A_{\mu} Q\;.
\end{equation}

The Feynman rule for the coupling of two new scalars to a photon, $s_1 s_2 \gamma_{\mu}$, for $s_1$ with charge $Q$ and $s_2 = s_1^*$, is given by
\begin{equation}
ie C_{s_1 s_2 \gamma}(p_1 - p_2)_{\mu}\;,
\end{equation}
where $C_{s_1 s_2 \gamma} = Q$ is the electric charge of scalar $s_1$.

The Feynman rule for the coupling of two new scalar to a $Z$ boson, $s_1 s_2 Z_{\mu}$, is given by
\begin{equation}
i e C_{s_1 s_2 Z} (p_1 - p_2)_{\mu}\;,
\end{equation}
where
\begin{align}
C_{\zr \zi Z} &= \frac{i}{2 s_Wc_W}\;,\nonumber\\
C_{\HkQ{1}{Q} \HkQ{1}{-Q} Z} &= \frac{1}{s_Wc_W} \left[ \left(Q-\frac{1}{2}\right) \cos^2\alpha_Q + \left(Q+\frac{1}{2}\right) \sin^2 \alpha_Q - Q s_W^2 \right]\;,\nonumber\\
C_{\HkQ{2}{Q} \HkQ{2}{-Q} Z} &= \frac{1}{s_Wc_W} \left[ \left(Q-\frac{1}{2}\right) \sin^2\alpha_Q + \left(Q+\frac{1}{2}\right) \cos^2 \alpha_Q - Q s_W^2 \right]\;,\nonumber\\
C_{\HkQ{1}{Q} \HkQ{2}{-Q} Z} = C_{\HkQ{2}{Q} \HkQ{1}{-Q} Z} &= \frac{1}{s_W c_W}\sin{\alpha_Q}\cos{\alpha_Q}\;,\nonumber\\
C_{\zp{\frac{n}{2}} \zeta^{-\frac{n}{2}} Z} &= \frac{1}{s_W c_W}\left[\frac{n-1}{2}-\frac{n}{2}s_W^2\right].
\end{align}
Note that the diagonal couplings $C_{\zr \zr Z} = C_{\zi \zi Z} = 0$ due to parity conservation.

The Feynman rule for the coupling of two new scalars to a $W$ boson, $s_1 s_2 W^{\pm}_{\mu}$, is given by
\begin{equation}
i e C_{s_1 s_2 W^{\pm}} (p_1 - p_2)_{\mu}\;.
\end{equation}
For compactness, we define the following coefficients for a given value of $n$:
\begin{equation}\begin{aligned}
T^+_Q &= \frac{1}{2}\sqrt{n^2 - 4 Q^2}\;,\\
T^-_Q &= \frac{1}{2}\sqrt{n^2 - 4 (Q-1)^2}\;.
\end{aligned}\end{equation}
Then the couplings of two scalars to $W^+$ are given by 
\begin{align}
C_{\zr H_1^- W^+} &= \frac{1}{2 s_W}\left[ \frac{n}{2} \cos\alpha_1 - T_{-1}^+ \sin\alpha_1 \right]\;,\nonumber\\
C_{\zr H_2^- W^+} &= \frac{1}{2 s_W}\left[ -\frac{n}{2} \sin\alpha_1 - T_{-1}^+ \cos\alpha_1 \right]\;,\nonumber\\
C_{\zi H_1^- W^+} &= \frac{i}{2 s_W} \left[ \frac{n}{2} \cos\alpha_1 + T_{-1}^+ \sin\alpha_1 \right]\;,\nonumber\\
C_{\zi H_2^- W^+} &= \frac{i}{2 s_W} \left[ -\frac{n}{2} \sin\alpha_1 + T_{-1}^+ \cos\alpha_1 \right]\;,\nonumber\\
C_{H_1^Q H_1^{-(Q+1)} W^+} &= \frac{1}{\sqrt{2} s_W} \left[ T_Q^+ \cos\alpha_Q \cos\alpha_{Q+1} - T_{-Q-1}^+ \sin\alpha_Q \sin\alpha_{Q+1} \right]\;,\nonumber\\
C_{H_1^Q H_2^{-(Q+1)} W^+} &= \frac{1}{\sqrt{2} s_W} \left[ -T_Q^+ \cos\alpha_Q \sin\alpha_{Q+1} - T_{-Q-1}^+ \sin\alpha_Q \cos\alpha_{Q+1} \right]\;,\nonumber\\
C_{H_2^Q H_1^{-(Q+1)} W^+} &= \frac{1}{\sqrt{2} s_W} \left[ -T_Q^+ \sin\alpha_Q \cos\alpha_{Q+1} - T_{-Q-1}^+ \cos\alpha_Q \sin\alpha_{Q+1} \right]\;,\nonumber\\
C_{H_2^Q H_2^{-(Q+1)} W^+} &= \frac{1}{\sqrt{2} s_W} \left[ T_Q^+ \sin\alpha_Q \sin\alpha_{Q+1} - T_{-Q-1}^+ \cos\alpha_Q \cos\alpha_{Q+1} \right]\;,\nonumber\\
C_{H_1^{\frac{n}{2}-1} \zeta^{-\frac{n}{2}} W^+} &= \frac{1}{\sqrt{2} s_W} T^+_{\frac{n}{2}-1} \cos\alpha_{\frac{n}{2}-1}\;,\nonumber\\
C_{H_2^{\frac{n}{2}-1} \zeta^{-\frac{n}{2}} W^+} &= -\frac{1}{\sqrt{2} s_W} T^+_{\frac{n}{2}-1} \sin\alpha_{\frac{n}{2}-1}.
\end{align}
The couplings of two scalars to $W^-$ are obtained using the relation
\begin{equation}
C_{s_2^* s_1^* W^-} = (C_{s_1 s_2 W^+})\co\;.
\end{equation}
Note that all the couplings $C_{s_1 s_2 W^+}$ are real except for those that involve one $\zi$, which are imaginary.

\section{One-loop RGEs}\label{sec:RGEs}

In this appendix, we provide the full one-loop renormalization group equations (RGEs) for the large multiplet models. We give the expressions for the mass parameters, $\mu_h^2$ and $M^2$, as well as for the gauge couplings, $g_i$, and the scalar quartic couplings, $\lambda_i$. The one-loop beta functions for various combinations of isospin and hypercharge were calculated up to $n=7$ in Ref.~\cite{Hamada:2015bra}. Here we present the expression for our $Y_\mult = 1$, $n=6$, $8$ models in our parameterization. The full scalar potential is given by
\begin{equation}\label{eq:theory_full_potential}
\begin{aligned}
V(\Phi,\,\mult) &= m^2 \Phi\ct \Phi + \lambda_1 \left(\Phi\ct \Phi\right)^2\\ &+ M^2 \mult\ct \mult + \lambda_2 \Phi\ct \Phi \mult\ct \mult + \lambda_3 \Phi\ct T_\Phi^a \Phi \mult\ct T_\mult^a \mult\\ &+ \left(\lambda_4 \cm{\Phi}\ct T_\Phi^a \Phi \mult\ct T_\mult^a \cm{\mult} + \mbox{h.c.}\right) + \left(\lambda_5\,[\Phi \mult]_3\ct\,[\mult \mult]_3 + \mbox{h.c.}\right)\\ &+ \lambda_6\,[\mult \mult]_{1}\ct\,[\mult \mult]_{1} + \lambda_7\,[\mult \mult]_{3}\ct\,[\mult \mult]_{3}\\ &+ \lambda_8\,[\mult \mult]_{5}\ct\,[\mult \mult]_{5} + \delta_{n8}\,\lambda_9\,[\mult \mult]_{7}\ct\,[\mult \mult]_{7}\;,
\end{aligned}
\end{equation}
where the terms $[\mult \mult]_{T}$ are the properly-normalized isospin-$T$ combinations of two $\mult$ fields, given in~\cref{sec:iso_combos}. The potential is organized in this manner to make manifest the counting of independent terms. Each valid combination of isospin and hypercharge is present. Recall that the large multiplet carries total isospin $T = \frac{n-1}{2}$ and hypercharge $Y_\mult = 1$. Even-isospin combinations are zero because they would be odd under interchange of two identical $\mult$ fields.
We will set $\lambda_5 = 0$ in order to preserve the global $Z_2$ symmetry $\mult \to -\mult$; hence we did not compute the RGE for $\lambda_5$.

The expression for the $U(1)_Y$ gauge coupling in a large multiplet model, $g_1$, is given by~\cite{Hamada:2015bra},
\begin{equation}
(4\pi)^2\,\frac{d g_1}{dt} = \left[\frac{41}{10} + \frac{1}{20}\,n\,Y_\mult^2\right]g_1^3\;,
\end{equation}
where $n$ is the size of the multiplet and $Y_\mult$ is the multiplet's hypercharge. We use the GUT normalization such that $g^{\prime} = \sqrt{\frac{3}{5}}g_1$, $g = g_2$, and $g_s = g_3$.

The expression for the $SU(2)_L$ gauge coupling in a large multiplet model, $g_2$ is given by~\cite{Logan:2015xpa},
\begin{equation}
(4\pi)^2\,\frac{d g_2}{dt} = -\left[\frac{19}{6} - \frac{n(n^2 - 1)}{36}\right]g_2^3\;,
\end{equation}
where $n$ is the size of the multiplet and $Y_i$ is the multiplet's hypercharge.

The expression for the $SU(3)_c$ gauge coupling, $g_3$ is unchanged by the presence of a large $SU(2)_L \otimes U(1)_Y$ multiplet, and is the same as in the SM, given by~\cite{Peskin:1995ev},
\begin{equation}
(4\pi)^2\,\frac{d g_3}{dt} = -\left[11 - \frac{2}{3}n_f\right]g_3^3\;,
\end{equation}
where $n_f$ is the number of coloured fermions (in our case, $n_f = 6$).

\subsection{$n = 6$ model}

\begin{equation}
16\pi^2\frac{dg_1}{dt} = \frac{22}{5}g_1^3\;,
\end{equation}
\begin{equation}
16\pi^2\frac{dg_2}{dt} = \frac{8}{3}g_2^3\;,
\end{equation}
\begin{equation}
16\pi^2\frac{dg_3}{dt} = -7g_3^3\;,
\end{equation}
\begin{equation}
16\pi^2\frac{dm^2}{dt} = - \frac{9}{10}g_1^2 m^2 - \frac{9}{2}g_2^2 m^2 + 12\lambda_1 m^2 + 12\lambda_2M^2 + 6y_t^2 m^2 + 6y_b^2 m^2 + 2y_{\tau}^2 m^2\;,
\end{equation}
\begin{equation}
16\pi^2\frac{dM^2}{dt} = -\frac{9}{10}g_1^2M^2 - \frac{105}{2}g_2^2M^2 + 4\lambda_2 m^2 + 4\lambda_6M^2 + \frac{28}{3}\lambda_7M^2 + \frac{44}{3}\lambda_8M^2\;,
\end{equation}
\begin{equation}\begin{aligned}
16\pi^2\frac{d\lambda_1}{dt} &= \frac{27}{200}g_1^4 + \frac{9}{8}g_2^4 + \frac{9}{20}g_1^2g_2^2 - \frac{9}{5}g_1^2\lambda_1 - 9g_2^2\lambda_1 + 24\lambda_1^2 + 6\lambda_2^2 + \frac{35}{8}\lambda_3^2 + 35\lambda_4^2\\&\hspace*{2em} + 12y_t^2\lambda_1 + 12y_b^2\lambda_1 + 4y_{\tau}^2\lambda_1 - 6y_t^4 - 6y_b^4 - 2y_{\tau}^4\;,
\end{aligned}\end{equation}
\begin{equation}\begin{aligned}
16\pi^2\frac{d\lambda_2}{dt} &= \frac{27}{100}g_1^4 + \frac{105}{4}g_2^4 - \frac{9}{5}g_1^2\lambda_2 - 57g_2^2\lambda_2 + 4\lambda_2^2 + \frac{35}{4}\lambda_3^2 + 70\lambda_4^2\\&\hspace*{2em} + 12\lambda_1\lambda_2 + 4\lambda_2\lambda_6 + \frac{28}{3}\lambda_2\lambda_7 + \frac{44}{3}\lambda_2\lambda_8 + 6y_t^2\lambda_2 + 6y_b^2\lambda_2 + 2y_{\tau}^2\lambda_2\;,
\end{aligned}\end{equation}
\begin{equation}\begin{aligned}
16\pi^2\frac{d\lambda_3}{dt} &= \frac{18}{5}g_1^2g_2^2 - \frac{9}{5}g_1^2\lambda_3 - 57g_2^2\lambda_3 + 16\lambda_4^2 + 4\lambda_1\lambda_3 + 8\lambda_2\lambda_3\\&\hspace*{2em} - \frac{124}{35}\lambda_3\lambda_6 - \frac{44}{15}\lambda_3\lambda_7 + \frac{220}{21}\lambda_3\lambda_8 + 6y_t^2\lambda_3 + 6y_b^2\lambda_3 + 2y_{\tau}^2\lambda_3\;,
\end{aligned}\end{equation}
\begin{equation}
16\pi^2\frac{d\lambda_4}{dt} = - \frac{9}{5}g_1^2\lambda_4 - 57g_2^2\lambda_4 + 4\lambda_1\lambda_4 + 8\lambda_2\lambda_4 + 4\lambda_3\lambda_4 + 4\lambda_4\lambda_6 + 6\lambda_4y_t^2 + 6\lambda_4y_b^2 + 2\lambda_4y_{\tau}^2\;,
\end{equation}
\begin{equation}\begin{aligned}
16\pi^2\frac{d\lambda_6}{dt} &= \frac{27}{200}g_1^4 + \frac{2697}{8}g_2^4 - \frac{279}{20}g_1^2g_2^2 -\frac{9}{5}g_1^2\lambda_6 - 105g_2^2\lambda_6 + 2\lambda_2^2 - \frac{31}{8}\lambda_3^2 + 35\lambda_4^2\\&\hspace*{2em} + \frac{8248}{1225}\lambda_6^2 + \frac{2054}{675}\lambda_7^2 + \frac{374}{1323}\lambda_8^2 + \frac{88}{75}\lambda_6\lambda_7 + \frac{2552}{147}\lambda_6\lambda_8 + \frac{308}{27}\lambda_7\lambda_8\;,
\end{aligned}\end{equation}
\begin{equation}\begin{aligned}
16\pi^2\frac{d\lambda_7}{dt} &= \frac{27}{200}g_1^4 + \frac{297}{8}g_2^4 - \frac{99}{20}g_1^2g_2^2 -\frac{9}{5}g_1^2\lambda_7 - 105g_2^2\lambda_7 + 2\lambda_2^2 - \frac{11}{8}\lambda_3^2\\&\hspace*{2em} + \frac{44}{175}\lambda_6^2 + \frac{21152}{2025}\lambda_7^2 + \frac{1496}{567}\lambda_8^2 + \frac{4108}{1575}\lambda_6\lambda_7 + \frac{44}{9}\lambda_6\lambda_8 + \frac{10868}{567}\lambda_7\lambda_8\;,
\end{aligned}\end{equation}
\begin{equation}\begin{aligned}
16\pi^2\frac{d\lambda_8}{dt} &= \frac{27}{200}g_1^4 + \frac{2025}{8}g_2^4 + \frac{45}{4}g_1^2g_2^2 -\frac{9}{5}g_1^2\lambda_8 - 105g_2^2\lambda_8 + 2\lambda_2^2 + \frac{25}{8}\lambda_3^2 + \frac{116}{49}\lambda_6^2\\&\hspace*{2em} + \frac{494}{81}\lambda_7^2 + \frac{98870}{3969}\lambda_8^2 + \frac{28}{9}\lambda_6\lambda_7 + \frac{68}{441}\lambda_6\lambda_8 + \frac{272}{81}\lambda_7\lambda_8\;.
\end{aligned}\end{equation}

\subsection{$n = 8$ model}

\begin{equation}
16\pi^2\frac{dg_1}{dt} = \frac{9}{2}g_1^3\;,
\end{equation}
\begin{equation}
16\pi^2\frac{dg_2}{dt} = \frac{65}{3}g_2^3\;,
\end{equation}
\begin{equation}
16\pi^2\frac{dg_3}{dt} = -7g_3^3\;,
\end{equation}
\begin{equation}
16\pi^2\frac{d m^2}{dt} = - \frac{9}{10}g_1^2 m^2 - \frac{9}{2}g_2^2 m^2 + 12\lambda_1 m^2 + 16\lambda_2M^2 + 6y_t^2 m^2 + 6y_b^2 m^2 + 2y_{\tau}^2 m^2\;,
\end{equation}
\begin{equation}
16\pi^2\frac{dM^2}{dt} = -\frac{9}{10}g_1^2M^2 - \frac{189}{2}g_2^2M^2 + 4\lambda_2 m^2 + 3\lambda_6M^2 + 7\lambda_7M^2 + 11\lambda_8M^2 + 15\lambda_9M^2\;,
\end{equation}
\begin{equation}\begin{aligned}
16\pi^2\frac{d\lambda_1}{dt} &= \frac{27}{200}g_1^4 + \frac{9}{8}g_2^4 + \frac{9}{20}g_1^2g_2^2 - \frac{9}{5}g_1^2\lambda_1 - 9g_2^2\lambda_1 + 24\lambda_1^2 + 8\lambda_2^2 + \frac{21}{2}\lambda_3^2 + 84\lambda_4^2\\&\hspace*{2em} + 12y_t^2\lambda_1 + 12y_b^2\lambda_1 + 4y_{\tau}^2\lambda_1 - 6y_t^4 - 6y_b^4 - 2y_{\tau}^4\;,
\end{aligned}\end{equation}
\begin{equation}\begin{aligned}
16\pi^2\frac{d\lambda_2}{dt} &= \frac{27}{100}g_1^4 + \frac{189}{4}g_2^4 - \frac{9}{5}g_1^2\lambda_2 - 99g_2^2\lambda_2 + 4\lambda_2^2 + \frac{63}{4}\lambda_3^2 + 126\lambda_4^2 + 12\lambda_1\lambda_2\\&\hspace*{2em} + 3\lambda_2\lambda_6 + 7\lambda_2\lambda_7 + 11\lambda_2\lambda_8 + 15\lambda_2\lambda_9 + 6y_t^2\lambda_2 + 6y_b^2\lambda_2 + 2y_{\tau}^2\lambda_2\;,
\end{aligned}\end{equation}
\begin{equation}\begin{aligned}
16\pi^2\frac{d\lambda_3}{dt} &= \frac{18}{5}g_1^2g_2^2 - \frac{9}{5}g_1^2\lambda_3 - 99g_2^2\lambda_3 + 16\lambda_4^2 + 4\lambda_1\lambda_3 + 8\lambda_2\lambda_3\\&\hspace*{2em} - \frac{59}{21}\lambda_3\lambda_6 - \frac{13}{3}\lambda_3\lambda_7 - \frac{11}{21}\lambda_3\lambda_8 + \frac{35}{3}\lambda_3\lambda_9 + 6y_t^2\lambda_3 + 6y_b^2\lambda_3 + 2y_{\tau}^2\lambda_3\;,
\end{aligned}\end{equation}
\begin{equation}\begin{aligned}
16\pi^2\frac{d\lambda_4}{dt} = - \frac{9}{5}g_1^2\lambda_4 - 99g_2^2\lambda_4 + 4\lambda_1\lambda_4 + 8\lambda_2\lambda_4 + 4\lambda_3\lambda_4 + 4\lambda_4\lambda_6 + 6y_t^2\lambda_4 + 6y_b^2\lambda_4 + 2y_{\tau}^2\lambda_4\;,
\end{aligned}\end{equation}
\begin{equation}\begin{aligned}
16\pi^2\frac{d\lambda_6}{dt} &= \frac{27}{200}g_1^4 + \frac{10089}{8}g_2^4 - \frac{531}{20}g_1^2g_2^2 - \frac{9}{5}g_1^2\lambda_6 - 189g_2^2\lambda_6 + 2\lambda_2^2 - \frac{59}{8}\lambda_3^2 + 84\lambda_4^2\\&\hspace*{2em} + \frac{10301}{1764}\lambda_6^2 - \frac{155}{132}\lambda_7^2 + \frac{25289}{2548}\lambda_8^2 + \frac{115}{5148}\lambda_9^2 + \frac{65}{18}\lambda_6\lambda_7 - \frac{11}{882}\lambda_6\lambda_8 - \frac{1}{6}\lambda_7\lambda_8\\&\hspace*{2em} + \frac{265}{18}\lambda_6\lambda_9 + \frac{2555}{198}\lambda_7\lambda_9 + \frac{545}{234}\lambda_8\lambda_9\;,
\end{aligned}\end{equation}
\begin{equation}\begin{aligned}
16\pi^2\frac{d\lambda_7}{dt} &= \frac{27}{200}g_1^4 + \frac{4329}{8}g_2^4 - \frac{351}{20}g_1^2g_2^2 - \frac{9}{5}g_1^2\lambda_7 - 189g_2^2\lambda_7 + 2\lambda_2^2\\&\hspace*{2em} - \frac{39}{8}\lambda_3^2 + \frac{65}{84}\lambda_6^2 + \frac{10063}{1452}\lambda_7^2 + \frac{1501}{1092}\lambda_8^2 + \frac{4715}{18876}\lambda_9^2 - \frac{155}{154}\lambda_6\lambda_7 - \frac{1}{14}\lambda_6\lambda_8\\&\hspace*{2em} + \frac{4735}{462}\lambda_7\lambda_8 + \frac{365}{66}\lambda_6\lambda_9 + \frac{3605}{242}\lambda_7\lambda_9 + \frac{2595}{286}\lambda_8\lambda_9\;,
\end{aligned}\end{equation}
\begin{equation}\begin{aligned}
16\pi^2\frac{d\lambda_8}{dt} &= \frac{27}{200}g_1^4 + \frac{9}{8}g_2^4 - \frac{27}{20}g_1^2g_2^2 - \frac{9}{5}g_1^2\lambda_8 - 189g_2^2\lambda_8 + 2\lambda_2^2 - \frac{3}{8}\lambda_3^2 - \frac{1}{588}\lambda_6^2\\&\hspace*{2em} + \frac{4735}{1452}\lambda_7^2 + \frac{1009231}{99372}\lambda_8^2 + \frac{621575}{245388}\lambda_9^2 - \frac{1}{22}\lambda_6\lambda_7 + \frac{6897}{1274}\lambda_6\lambda_8 + \frac{1501}{858}\lambda_7\lambda_8\\&\hspace*{2em} + \frac{545}{858}\lambda_6\lambda_9 + \frac{18165}{3146}\lambda_7\lambda_9 + \frac{68875}{3718}\lambda_8\lambda_9\;,
\end{aligned}\end{equation}
\begin{equation}\begin{aligned}
16\pi^2\frac{d\lambda_9}{dt} &= \frac{27}{200}g_1^4 + \frac{7497}{8}g_2^4 + \frac{441}{20}g_1^2g_2^2 - \frac{9}{5}g_1^2\lambda_9 - 189g_2^2\lambda_9 + 2\lambda_2^2 + \frac{49}{8}\lambda_3^2 + \frac{53}{36}\lambda_6^2\\&\hspace*{2em} + \frac{5047}{1452}\lambda_7^2 + \frac{13775}{2028}\lambda_8^2 + \frac{18419861}{736164}\lambda_9^2 + \frac{511}{198}\lambda_6\lambda_7 + \frac{109}{234}\lambda_6\lambda_8 + \frac{1211}{286}\lambda_7\lambda_8\\&\hspace*{2em} + \frac{23}{2574}\lambda_6\lambda_9 + \frac{6601}{28314}\lambda_7\lambda_9 + \frac{124315}{33462}\lambda_8\lambda_9\;.
\end{aligned}\end{equation}

\section{Isospin combinations}\label{sec:iso_combos}

We can combine two isospin multiplets with total isospin $T_{(1)}$ and $T_{(2)}$, and third component of isospin $T_{(1)}^3$ and $T_{(2)}^3$, respectively, into a multiplet with total isospin $T$,
\begin{equation}\label{eq:app_isocombo_ket}
\left|T,\,T^3\right> = \left|T_{(1)},\,T^{3}_{(1)}\right> \otimes \left|T_{(2)},\,T^{3}_{(2)}\right>
\end{equation}
where $T^3 = T_{(1)}^3 + T_{(2)}^3$ and $T = T_{(1)} \oplus T_{(2)} \in [-(|T_{(1)}| + |T_{(2)}|),\,|T_{(1)}| + |T_{(2)}|]$. To combine these, we act with the usual ladder operators,
\begin{equation}\begin{aligned}
J^{+}\left|j,\,m\right> &= \sqrt{j(j+1) - m(m+1)}\left|j,\,m+1\right>\;,\\
J^{-}\left|j,\,m\right> &= \sqrt{j(j+1) - m(m-1)}\left|j,\,m-1\right>\;,
\end{aligned}\end{equation}
on the states of~\cref{eq:app_isocombo_ket}.

To clarify some notation, many of the expressions for the field combinations are quite long and will not fit on one line. To simplify presentation, these components will be displayed separately, and so we define the combination with total isospin $T$ as
\begin{equation}
{[\mult \mult]}_{T} \equiv \left(\begin{matrix}{[\mult \mult]}_{T}^{(T)}\\ {[\mult \mult]}_{T}^{(T-1)}\\ \vdots\\ {[\mult \mult]}_{T}^{(-T+1)}\\ {[\mult \mult]}_{T}^{(-T)}\end{matrix}\right)\;.
\end{equation}

\subsection{$n=6$ model}

The combination $[\Phi \mult]_3$ is
\begin{equation}
{[\Phi \mult]}_{3} = \left(
\begin{array}{c}
\zeta^{+3}\,\phi^{+}\\
\sqrt{\frac{5}{6}}\,\zeta^{+2}\,\phi^{+} + \frac{1}{\sqrt{6}}\,\zeta^{+3}\,\phi^{0}\\
\sqrt{\frac{2}{3}}\,\zeta^{+}\,\phi^{+} + \frac{1}{\sqrt{3}}\,\zeta^{+2}\,\phi^{0}\\
\frac{1}{\sqrt{2}}\,\zeta^{0}\,\phi^{+} + \frac{1}{\sqrt{2}}\,\zeta^{+1}\,\phi^{0}\\
\frac{1}{\sqrt{3}}\,\zeta^{-1}\,\phi^{+} + \sqrt{\frac{2}{3}}\,\zeta^{0}\,\phi^{0}\\
\frac{1}{\sqrt{6}}\,\zeta^{-2}\,\phi^{+} + \sqrt{\frac{5}{6}}\,\zeta^{-1}\,\phi^{0}\\
\zeta^{-2}\,\phi^{0}
\end{array}
\right)\;.
\end{equation}

The combination $[\mult \mult]_1$ is
\begin{align}
\nonumber{[\mult \mult]}_{1}^{(1)} &= \frac{1}{\sqrt{7}}\,\zeta^{-}\,\zeta^{+3} - \sqrt{\frac{8}{35}}\,\zeta^{0}\,\zeta^{+2} + \frac{3}{\sqrt{35}}\,\zeta^{+}\,\zeta^{+} - \sqrt{\frac{8}{35}}\,\zeta^{+2}\,\zeta^{0} + \frac{1}{\sqrt{7}}\zeta^{+3}\,\zeta^{-}\;,\\
\nonumber{[\mult \mult]}_{1}^{(0)} &= \sqrt{\frac{5}{14}}\,\zeta^{-2}\,\zeta^{+3} - \frac{3}{\sqrt{70}}\,\zeta^{-}\,\zeta^{+2} + \frac{1}{\sqrt{70}}\,\zeta^{0}\,\zeta^{+} + \frac{1}{\sqrt{70}}\zeta^{+}\,\zeta^{0}\\\nonumber&\hspace*{1em} - \frac{3}{\sqrt{70}}\,\zeta^{+2}\,\zeta^{-} + \sqrt{\frac{5}{14}}\,\zeta^{+3}\,\zeta^{-2}\;,\\
{[\mult \mult]}_{1}^{(-1)} &= \frac{1}{\sqrt{7}}\zeta^{-2}\,\zeta^{+2} - \sqrt{\frac{8}{35}}\,\zeta^{-}\,\zeta^{+} + \frac{3}{\sqrt{35}}\,\zeta^{0}\,\zeta^{0} - \sqrt{\frac{8}{35}}\,\zeta^{+}\,\zeta^{-} + \frac{1}{\sqrt{7}}\,\zeta^{+2}\,\zeta^{-2}\;.
\end{align}

The combination $[\mult \mult]_3$ is
\begin{align}
\nonumber{[\mult \mult]}_{3}^{(3)} &= \sqrt{\frac{5}{18}}\,\zeta^{+}\,\zeta^{+3} - \frac{2}{3}\,\zeta^{+2}\,\zeta^{+2} + \sqrt{\frac{5}{18}}\,\zeta^{+3}\,\zeta^{+}\;,\\
\nonumber{[\mult \mult]}_{3}^{(2)} &= \sqrt{\frac{5}{12}}\,\zeta^{0}\,\zeta^{+3} - \frac{1}{\sqrt{12}}\,\zeta^{+}\,\zeta^{+2} - \frac{1}{\sqrt{12}}\,\zeta^{+2}\,\zeta^{+} + \sqrt{\frac{5}{12}}\,\zeta^{+3}\,\zeta^{0}\;,\\
\nonumber{[\mult \mult]}_{3}^{(1)} &= \frac{1}{\sqrt{3}}\,\zeta^{-}\,\zeta^{+3} + \frac{1}{\sqrt{30}}\,\zeta^{0}\,\zeta^{+2} - \frac{2}{\sqrt{15}}\,\zeta^{+}\,\zeta^{+} + \frac{1}{\sqrt{30}}\,\zeta^{+2}\,\zeta^{0} + \frac{1}{\sqrt{3}}\,\zeta^{+3}\,\zeta^{-}\;,\\
\nonumber{[\mult \mult]}_{3}^{(0)} &= \frac{\sqrt{5}}{6}\,\zeta^{-2}\,\zeta^{+3} + \frac{7}{6 \sqrt{5}}\,\zeta^{-}\,\zeta^{+2} - \frac{2}{3 \sqrt{5}}\,\zeta^{0}\,\zeta^{+} - \frac{2}{3 \sqrt{5}}\,\zeta^{+}\,\zeta^{0}\\\nonumber&\hspace*{1em} + \frac{7}{6 \sqrt{5}}\,\zeta^{+2}\,\zeta^{-} + \frac{\sqrt{5}}{6}\,\zeta^{+3}\,\zeta^{-2}\;,\\
\nonumber{[\mult \mult]}_{3}^{(-1)} &= \frac{1}{\sqrt{3}}\,\zeta^{-2}\,\zeta^{+2} + \frac{1}{\sqrt{30}}\,\zeta^{-}\,\zeta^{+} - \frac{2}{\sqrt{15}}\,\zeta^{0}\,\zeta^{0} + \frac{1}{\sqrt{30}}\,\zeta^{+}\,\zeta^{-} + \frac{1}{\sqrt{3}}\,\zeta^{+2}\,\zeta^{-2}\;,\\
\nonumber{[\mult \mult]}_{3}^{(-2)} &= \sqrt{\frac{5}{12}}\,\zeta^{-2}\,\zeta^{+} - \frac{1}{\sqrt{12}}\,\zeta^{-}\,\zeta^{0} - \frac{1}{\sqrt{12}}\,\zeta^{0}\,\zeta^{-} + \sqrt{\frac{5}{12}}\,\zeta^{+}\,\zeta^{-2}\;,\\
{[\mult \mult]}_{3}^{(-3)} &= \sqrt{\frac{5}{18}}\,\zeta^{-2}\,\zeta^{0} - \frac{2}{3}\,\zeta^{-}\,\zeta^{-} + \sqrt{\frac{5}{18}}\,\zeta^{0}\,\zeta^{-2}\;.
\end{align}

The combination $[\mult \mult]_5$ is
\begin{align}
\nonumber{[\mult \mult]}_{5}^{(5)} &= \zeta^{+3}\,\zeta^{+3}\;,\\
\nonumber{[\mult \mult]}_{5}^{(4)} &= \frac{1}{\sqrt{2}}\,\zeta^{+2}\,\zeta^{+3} + \frac{1}{\sqrt{2}}\,\zeta^{+3}\,\zeta^{+2}\;,\\
\nonumber{[\mult \mult]}_{5}^{(3)} &= \frac{\sqrt{2}}{3} \zeta^{+}\,\zeta^{+3} + \frac{\sqrt{5}}{3} \zeta^{+2}\,\zeta^{+2} + \frac{\sqrt{2}}{3} \zeta^{+3}\,\zeta^{+}\;,\\
\nonumber{[\mult \mult]}_{5}^{(2)} &= \frac{1}{2 \sqrt{3}}\,\zeta^{0}\,\zeta^{+3} + \frac{1}{2} \sqrt{\frac{5}{3}} \zeta^{+}\,\zeta^{+2} + \frac{1}{2} \sqrt{\frac{5}{3}} \zeta^{+2}\,\zeta^{+} + \frac{1}{2 \sqrt{3}}\,\zeta^{+3}\,\zeta^{0}\;,\\
\nonumber{[\mult \mult]}_{5}^{(1)} &= \frac{1}{\sqrt{42}}\,\zeta^{-}\,\zeta^{+3} + \sqrt{\frac{5}{21}} \zeta^{0}\,\zeta^{+2} + \sqrt{\frac{10}{21}} \zeta^{+}\,\zeta^{+} + \sqrt{\frac{5}{21}} \zeta^{+2}\,\zeta^{0} + \frac{1}{\sqrt{42}}\,\zeta^{+3}\,\zeta^{-}\;,\\
\nonumber{[\mult \mult]}_{5}^{(0)} &= \frac{1}{6 \sqrt{7}}\,\zeta^{-2}\,\zeta^{+3} + \frac{5}{6 \sqrt{7}}\,\zeta^{-}\,\zeta^{+2} + \frac{5}{3 \sqrt{7}}\,\zeta^{0}\,\zeta^{+} + \frac{5}{3 \sqrt{7}}\,\zeta^{+}\,\zeta^{0}\\\nonumber&\hspace*{1em} + \frac{5}{6 \sqrt{7}}\,\zeta^{+2}\,\zeta^{-} + \frac{1}{6 \sqrt{7}}\,\zeta^{+3}\,\zeta^{-2}\;,\\
\nonumber{[\mult \mult]}_{5}^{(-1)} &= \frac{1}{\sqrt{42}}\,\zeta^{-2}\,\zeta^{+2} + \sqrt{\frac{5}{21}} \zeta^{-}\,\zeta^{+} + \sqrt{\frac{10}{21}} \zeta^{0}\,\zeta^{0} + \sqrt{\frac{5}{21}}\,\zeta^{+}\,\zeta^{-} + \frac{1}{\sqrt{42}}\,\zeta^{+2}\,\zeta^{-2}\;,\\
\nonumber{[\mult \mult]}_{5}^{(-2)} &= \frac{1}{2 \sqrt{3}}\,\zeta^{-2}\,\zeta^{+} + \frac{1}{2} \sqrt{\frac{5}{3}} \zeta^{-}\,\zeta^{0} + \frac{1}{2} \sqrt{\frac{5}{3}}\,\zeta^{0}\,\zeta^{-} + \frac{1}{2 \sqrt{3}}\,\zeta^{+}\,\zeta^{-2}\;,\\
\nonumber{[\mult \mult]}_{5}^{(-3)} &= \frac{\sqrt{2}}{3} \zeta^{-2}\,\zeta^{0} + \frac{\sqrt{5}}{3}\,\zeta^{-}\,\zeta^{-} + \frac{\sqrt{2}}{3} \zeta^{0}\,\zeta^{-2}\;,\\
\nonumber{[\mult \mult]}_{5}^{(-4)} &= \frac{1}{\sqrt{2}}\,\zeta^{-2}\,\zeta^{-} + \frac{1}{\sqrt{2}}\,\zeta^{-}\,\zeta^{-2}\;,\\
{[\mult \mult]}_{5}^{(-5)} &= \zeta^{-2}\,\zeta^{-2}\;.
\end{align}

\subsection{$n=8$ model}

The combination $[\Phi \mult]_3$ is
\begin{equation}
[\Phi \mult]_{3} = \left(
\begin{array}{c}
\frac{1}{2} \sqrt{\frac{7}{2}} \zeta^{+4}\,\phi^{0} - \frac{1}{2 \sqrt{2}}\,\zeta^{+3}\,\phi^{+}\\
\frac{1}{2} \sqrt{3} \zeta^{+3}\,\phi^{0} - \frac{1}{2} \zeta^{+2}\,\phi^{+}\\
\frac{1}{2} \sqrt{\frac{5}{2}} \zeta^{+2}\,\phi^{0} - \frac{1}{2} \sqrt{\frac{3}{2}} \zeta^{+}\,\phi^{+}\\
\frac{1}{\sqrt{2}}\,\zeta^{+}\,\phi^{0} - \frac{1}{\sqrt{2}}\,\zeta^{0}\,\phi^{+}\\
\frac{1}{2} \sqrt{\frac{3}{2}} \zeta^{0}\,\phi^{0} - \frac{1}{2} \sqrt{\frac{5}{2}} \zeta^{-}\,\phi^{+}\\
\frac{1}{2} \zeta^{-}\,\phi^{0} - \frac{1}{2} \sqrt{3} \zeta^{-2}\,\phi^{+}\\
\frac{1}{2 \sqrt{2}}\zeta^{-2}\,\phi^{0} - \frac{1}{2} \sqrt{\frac{7}{2}} \zeta^{-3}\,\phi^{+}
\end{array}
\right)\;.
\end{equation}

The combination $[\mult \mult]_1$ is
\begin{align}
\nonumber{[\mult \mult]}_{1}^{(1)} &= \frac{1}{2\sqrt{3}}\,\zeta^{-2}\,\zeta^{+4} - \frac{1}{\sqrt{7}}\,\zeta^{-}\,\zeta^{+3} + \frac{1}{2}\sqrt{\frac{5}{7}}\,\zeta^{0}\,\zeta^{+2} - \frac{2}{\sqrt{21}}\,\zeta^{+}\,\zeta^{+}\\\nonumber&\hspace*{1em} + \frac{1}{2}\sqrt{\frac{5}{7}}\,\zeta^{+2}\,\zeta^{0} - \frac{1}{\sqrt{7}}\,\zeta^{+3}\,\zeta^{-} + \frac{1}{2\sqrt{3}}\,\zeta^{+4}\,\zeta^{-2}\;,\\
\nonumber{[\mult \mult]}_{1}^{(0)} &= \frac{1}{2}\sqrt{\frac{7}{6}}\,\zeta^{-3}\,\zeta^{+4} - \frac{5}{2\sqrt{42}}\,\zeta^{-2}\,\zeta^{+3} + \frac{1}{2}\sqrt{\frac{3}{14}}\,\zeta^{-}\,\zeta^{+2} - \frac{1}{2\sqrt{42}}\,\zeta^{0}\,\zeta^{+}\\\nonumber&\hspace*{1em} - \frac{1}{2\sqrt{42}}\,\zeta^{+}\,\zeta^{0} + \frac{1}{2}\sqrt{\frac{3}{14}}\,\zeta^{+2}\,\zeta^{-} - \frac{5}{2\sqrt{42}}\,\zeta^{+3}\,\zeta^{-2} + \frac{1}{2}\sqrt{\frac{7}{6}}\,\zeta^{+4}\,\zeta^{-3}\;,\\
{[\mult \mult]}_{1}^{(-1)} &= \frac{1}{2\sqrt{3}}\,\zeta^{-3}\,\zeta^{+3} - \frac{1}{\sqrt{7}}\,\zeta^{-2}\,\zeta^{+2} + \frac{1}{2}\sqrt{\frac{5}{7}}\,\zeta^{-}\,\zeta^{+} - \frac{2}{\sqrt{21}}\,\zeta^{0}\,\zeta^{0}\\\nonumber&\hspace*{1em} + \frac{1}{2}\sqrt{\frac{5}{7}}\,\zeta^{+}\,\zeta^{-} - \frac{1}{\sqrt{7}}\,\zeta^{+2}\,\zeta^{-2} + \frac{1}{2\sqrt{3}}\,\zeta^{+3}\,\zeta^{-3}\;.
\end{align}

The combination $[\mult \mult]_3$ is
\begin{align}
\nonumber{[\mult \mult]}_{3}^{(3)} &= \sqrt{\frac{7}{66}}\,\zeta^{0}\,\zeta^{+4} - 2\sqrt{\frac{2}{33}}\,\zeta^{+}\,\zeta^{+3} +\sqrt{\frac{10}{33}}\,\zeta^{+2}\,\zeta^{+2} - 2\sqrt{\frac{2}{33}}\,\zeta^{+3}\,\zeta^{+} + \sqrt{\frac{7}{66}}\,\zeta^{+4}\,\zeta^{0}\;,\\
\nonumber{[\mult \mult]}_{3}^{(2)} &= \frac{1}{2}\sqrt{\frac{35}{33}}\,\zeta^{-}\,\zeta^{+4} - \frac{3}{2\sqrt{11}}\,\zeta^{0}\,\zeta^{+3} + \frac{1}{\sqrt{33}}\,\zeta^{+}\,\zeta^{+2} + \frac{1}{\sqrt{33}}\,\zeta^{+2}\,\zeta^{+}\\\nonumber&\hspace*{1em} - \frac{3}{2\sqrt{11}}\,\zeta^{+3}\,\zeta^{0} + \frac{1}{2}\sqrt{\frac{35}{33}}\,\zeta^{+4}\,\zeta^{-}\;,\\
\nonumber{[\mult \mult]}_{3}^{(1)} &= \sqrt{\frac{7}{22}}\,\zeta^{-2}\,\zeta^{+4} - \frac{1}{\sqrt{66}}\,\zeta^{-}\,\zeta^{+3} - \sqrt{\frac{5}{66}}\,\zeta^{0}\,\zeta^{+2} + \sqrt{\frac{2}{11}}\,\zeta^{+}\,\zeta^{+}\\\nonumber&\hspace*{1em} - \sqrt{\frac{5}{66}}\,\zeta^{+2}\,\zeta^{0} - \frac{1}{\sqrt{66}}\,\zeta^{+3}\,\zeta^{-} + \sqrt{\frac{7}{22}}\,\zeta^{+4}\,\zeta^{-2}\;,\\
\nonumber{[\mult \mult]}_{3}^{(0)} &= \frac{7}{2\sqrt{66}}\,\zeta^{-3}\,\zeta^{+4} + \frac{5}{2\sqrt{66}}\,\zeta^{-2}\,\zeta^{+3} - \frac{7}{2\sqrt{66}}\,\zeta^{-}\,\zeta^{+2} + \frac{1}{2}\sqrt{\frac{3}{22}}\,\zeta^{0}\,\zeta^{+}\\\nonumber&\hspace*{1em} + \frac{1}{2}\sqrt{\frac{3}{22}}\,\zeta^{+}\,\zeta^{0} - \frac{7}{2\sqrt{66}}\,\zeta^{+2}\,\zeta^{-} + \frac{5}{2\sqrt{66}}\,\zeta^{+3}\,\zeta^{-2} + \frac{7}{2\sqrt{66}}\,\zeta^{+4}\,\zeta^{-3}\;,\\
\nonumber{[\mult \mult]}_{3}^{(-1)} &= \sqrt{\frac{7}{22}}\,\zeta^{-3}\,\zeta^{+3} - \frac{1}{\sqrt{66}}\,\zeta^{-2}\,\zeta^{+2} - \sqrt{\frac{5}{66}}\,\zeta^{-}\,\zeta^{+} + \sqrt{\frac{2}{11}}\,\zeta^{0}\,\zeta^{0}\\\nonumber&\hspace*{1em} - \sqrt{\frac{5}{66}}\,\zeta^{+}\,\zeta^{-} - \frac{1}{\sqrt{66}}\,\zeta^{+2}\,\zeta^{-2} + \sqrt{\frac{7}{22}}\,\zeta^{+3}\,\zeta^{-3}\;,\\
\nonumber{[\mult \mult]}_{3}^{(-2)} &= \frac{1}{2}\sqrt{\frac{35}{33}}\,\zeta^{-3}\,\zeta^{+2} - \frac{3}{2\sqrt{11}}\,\zeta^{-2}\,\zeta^{+} + \frac{1}{\sqrt{33}}\,\zeta^{-}\,\zeta^{0} + \frac{1}{\sqrt{33}}\,\zeta^{0}\,\zeta^{-}\\\nonumber&\hspace*{1em} - \frac{3}{2\sqrt{11}}\,\zeta^{+}\,\zeta^{-2} + \frac{1}{2}\sqrt{\frac{35}{33}}\,\zeta^{+2}\,\zeta^{-3}\;,\\
{[\mult \mult]}_{3}^{(-3)} &= \sqrt{\frac{7}{66}}\,\zeta^{-3}\,\zeta^{+} - 2\sqrt{\frac{2}{33}}\,\zeta^{-2}\,\zeta^{0}+\sqrt{\frac{10}{33}}\,\zeta^{-}\,\zeta^{-} - 2\sqrt{\frac{2}{33}}\,\zeta^{0}\,\zeta^{-2}+\sqrt{\frac{7}{66}}\,\zeta^{+}\,\zeta^{-3}\;.
\end{align}

The combination $[\mult \mult]_5$ is
\begin{align}
\nonumber{[\mult \mult]}_{5}^{(5)} &= \sqrt{\frac{7}{26}}\,\zeta^{+2}\,\zeta^{+4} - \sqrt{\frac{6}{13}}\,\zeta^{+3}\,\zeta^{+3} + \sqrt{\frac{7}{26}}\,\zeta^{+4}\,\zeta^{+2}\;,\\
\nonumber{[\mult \mult]}_{5}^{(4)} &= \frac{1}{2}\sqrt{\frac{21}{13}}\,\zeta^{+}\,\zeta^{+4} - \frac{1}{2}\sqrt{\frac{5}{13}}\,\zeta^{+2}\,\zeta^{+3} - \frac{1}{2}\sqrt{\frac{5}{13}}\,\zeta^{+3}\,\zeta^{+2} + \frac{1}{2}\sqrt{\frac{21}{13}}\,\zeta^{+4}\,\zeta^{+}\;,\\
\nonumber{[\mult \mult]}_{5}^{(3)} &= \sqrt{\frac{14}{39}}\,\zeta^{0}\,\zeta^{+4} + \frac{1}{\sqrt{78}}\,\zeta^{+}\,\zeta^{+3} - \sqrt{\frac{10}{39}}\,\zeta^{+2}\,\zeta^{+2} + \frac{1}{\sqrt{78}}\,\zeta^{+3}\,\zeta^{+} + \sqrt{\frac{14}{39}}\,\zeta^{+4}\,\zeta^{0}\;,\\
\nonumber{[\mult \mult]}_{5}^{(2)} &= \frac{1}{2}\sqrt{\frac{35}{39}}\,\zeta^{-}\,\zeta^{+4} + \frac{3}{2\sqrt{13}}\,\zeta^{0}\,\zeta^{+3} - \frac{2}{\sqrt{39}}\,\zeta^{+}\,\zeta^{+2} - \frac{2}{\sqrt{39}}\,\zeta^{+2}\,\zeta^{+}\\\nonumber&\hspace*{1em} + \frac{3}{2\sqrt{13}}\,\zeta^{+3}\,\zeta^{0} + \frac{1}{2}\sqrt{\frac{35}{39}}\,\zeta^{+4}\,\zeta^{-}\;,\\
\nonumber{[\mult \mult]}_{5}^{(1)} &= \frac{1}{2}\sqrt{\frac{5}{13}}\,\zeta^{-2}\,\zeta^{+4} + 4\sqrt{\frac{5}{273}}\,\zeta^{-}\,\zeta^{+3} + \frac{1}{2\sqrt{273}}\,\zeta^{0}\,\zeta^{+2} - 2\sqrt{\frac{5}{91}}\,\zeta^{+}\,\zeta^{+}\\\nonumber&\hspace*{1em} + \frac{1}{2\sqrt{273}}\,\zeta^{+2}\,\zeta^{0} + 4\sqrt{\frac{5}{273}}\,\zeta^{+3}\,\zeta^{-} + \frac{1}{2}\sqrt{\frac{5}{13}}\,\zeta^{+4}\,\zeta^{-2}\;,\\
\nonumber{[\mult \mult]}_{5}^{(0)} &= \frac{1}{2}\sqrt{\frac{7}{78}}\,\zeta^{-3}\,\zeta^{+4} + \frac{23}{2\sqrt{546}}\,\zeta^{-2}\,\zeta^{+3} + \frac{17}{2\sqrt{546}}\,\zeta^{-}\,\zeta^{+2} - \frac{5}{2}\sqrt{\frac{3}{182}}\,\zeta^{0}\,\zeta^{+}\\\nonumber&\hspace*{1em} - \frac{5}{2}\sqrt{\frac{3}{182}}\,\zeta^{+}\,\zeta^{0} + \frac{17}{2\sqrt{546}}\,\zeta^{+2}\,\zeta^{-} + \frac{23}{2\sqrt{546}}\,\zeta^{+3}\,\zeta^{-2} + \frac{1}{2}\sqrt{\frac{7}{78}}\,\zeta^{+4}\,\zeta^{-3}\;,\\
\nonumber{[\mult \mult]}_{5}^{(-1)} &= \frac{1}{2}\sqrt{\frac{5}{13}}\,\zeta^{-3}\,\zeta^{+3} + 4\sqrt{\frac{5}{273}}\,\zeta^{-2}\,\zeta^{+2} + \frac{1}{2\sqrt{273}}\,\zeta^{-}\,\zeta^{+} - 2\sqrt{\frac{5}{91}}\,\zeta^{0}\,\zeta^{0}\\\nonumber&\hspace*{1em} + \frac{1}{2\sqrt{273}}\,\zeta^{+}\,\zeta^{-} + 4\sqrt{\frac{5}{273}}\,\zeta^{+2}\,\zeta^{-2} + \frac{1}{2}\sqrt{\frac{5}{13}}\,\zeta^{+3}\,\zeta^{-3}\;,\\
\nonumber{[\mult \mult]}_{5}^{(-2)} &= \frac{1}{2}\sqrt{\frac{35}{39}}\,\zeta^{-3}\,\zeta^{+2} + \frac{3}{2\sqrt{13}}\,\zeta^{-2}\,\zeta^{+} - \frac{2}{\sqrt{39}}\,\zeta^{-}\,\zeta^{0} - \frac{2}{\sqrt{39}}\,\zeta^{0}\,\zeta^{-}\\\nonumber&\hspace*{1em} + \frac{3}{2\sqrt{13}}\,\zeta^{+}\,\zeta^{-2} + \frac{1}{2}\sqrt{\frac{35}{39}}\,\zeta^{+2}\,\zeta^{-3}\;,\\
\nonumber{[\mult \mult]}_{5}^{(-3)} &= \sqrt{\frac{14}{39}}\,\zeta^{-3}\,\zeta^{+} + \frac{1}{\sqrt{78}}\,\zeta^{-2}\,\zeta^{0} - \sqrt{\frac{10}{39}}\,\zeta^{-}\,\zeta^{-} + \frac{1}{\sqrt{78}}\,\zeta^{0}\,\zeta^{-2} + \sqrt{\frac{14}{39}}\,\zeta^{+}\,\zeta^{-3}\;,\\
\nonumber{[\mult \mult]}_{5}^{(-4)} &= \frac{1}{2}\sqrt{\frac{21}{13}}\,\zeta^{-3}\,\zeta^{0} - \frac{1}{2}\sqrt{\frac{5}{13}}\,\zeta^{-2}\,\zeta^{-} - \frac{1}{2}\sqrt{\frac{5}{13}}\,\zeta^{-}\,\zeta^{-2} + \frac{1}{2}\sqrt{\frac{21}{13}}\,\zeta^{0}\,\zeta^{-3}\;,\\
{[\mult \mult]}_{5}^{(-5)} &= \sqrt{\frac{7}{26}}\,\zeta^{-3}\,\zeta^{-} - \sqrt{\frac{6}{13}}\,\zeta^{-2}\,\zeta^{-2} + \sqrt{\frac{7}{26}}\,\zeta^{-}\,\zeta^{-3}\;.
\end{align}

The combination $[\mult \mult]_7$ is
\begin{align}
\nonumber{[\mult \mult]}_{7}^{(7)} &= \zeta^{+4}\,\zeta^{+4}\;,\\
\nonumber{[\mult \mult]}_{7}^{(6)} &= \frac{1}{\sqrt{2}}\,\zeta^{+3}\,\zeta^{+4} + \frac{1}{\sqrt{2}}\,\zeta^{+4}\,\zeta^{+3}\;,\\
\nonumber{[\mult \mult]}_{7}^{(5)} &= \sqrt{\frac{3}{13}}\,\zeta^{+2}\,\zeta^{+4} + \sqrt{\frac{7}{13}}\,\zeta^{+3}\,\zeta^{+3} + \sqrt{\frac{3}{13}}\,\zeta^{+4}\,\zeta^{+2}\;,\\
\nonumber{[\mult \mult]}_{7}^{(4)} &= \frac{1}{2}\sqrt{\frac{5}{13}}\,\zeta^{+}\,\zeta^{+4} + \frac{1}{2}\sqrt{\frac{21}{13}}\,\zeta^{+2}\,\zeta^{+3} + \frac{1}{2}\sqrt{\frac{21}{13}}\,\zeta^{+3}\,\zeta^{+2} + \frac{1}{2}\sqrt{\frac{5}{13}}\,\zeta^{+4}\,\zeta^{+}\;,\\
\nonumber{[\mult \mult]}_{7}^{(3)} &= \sqrt{\frac{5}{143}}\,\zeta^{0}\,\zeta^{+4} + \sqrt{\frac{35}{143}}\,\zeta^{+}\,\zeta^{+3} + \sqrt{\frac{63}{143}}\,\zeta^{+2}\,\zeta^{+2} + \sqrt{\frac{35}{143}}\,\zeta^{+3}\,\zeta^{+} + \sqrt{\frac{5}{143}}\,\zeta^{+4}\,\zeta^{0}\;,\\
\nonumber{[\mult \mult]}_{7}^{(2)} &= \sqrt{\frac{3}{286}}\,\zeta^{-}\,\zeta^{+4} + \sqrt{\frac{35}{286}}\,\zeta^{0}\,\zeta^{+3} + \sqrt{\frac{105}{286}}\,\zeta^{+}\,\zeta^{+2} + \sqrt{\frac{105}{286}}\,\zeta^{+2}\,\zeta^{+}\\\nonumber&\hspace*{1em} + \sqrt{\frac{35}{286}}\,\zeta^{+3}\,\zeta^{0} + \sqrt{\frac{3}{286}}\,\zeta^{+4}\,\zeta^{-}\;,\\
\nonumber{[\mult \mult]}_{7}^{(1)} &= \frac{1}{\sqrt{429}}\,\zeta^{-2}\,\zeta^{+4} + \sqrt{\frac{7}{143}}\,\zeta^{-}\,\zeta^{+3} + \sqrt{\frac{35}{143}}\,\zeta^{0}\,\zeta^{+2} + 5\sqrt{\frac{7}{429}}\,\zeta^{+}\,\zeta^{+}\\\nonumber&\hspace*{1em} + \sqrt{\frac{35}{143}}\,\zeta^{+2}\,\zeta^{0} + \sqrt{\frac{7}{143}}\,\zeta^{+3}\,\zeta^{-} + \frac{1}{\sqrt{429}}\,\zeta^{+4}\,\zeta^{-2}\;,\\
\nonumber{[\mult \mult]}_{7}^{(0)} &= \frac{1}{2\sqrt{858}}\,\zeta^{-3}\,\zeta^{+4} + \frac{7}{2\sqrt{858}}\,\zeta^{-2}\,\zeta^{+3} + \frac{7}{2}\sqrt{\frac{3}{286}}\,\zeta^{-}\,\zeta^{+2} + \frac{35}{2\sqrt{858}}\,\zeta^{0}\,\zeta^{+}\\\nonumber&\hspace*{1em} + \frac{35}{2\sqrt{858}}\,\zeta^{+}\,\zeta^{0} + \frac{7}{2}\sqrt{\frac{3}{286}}\,\zeta^{+2}\,\zeta^{-} + \frac{7}{2\sqrt{858}}\,\zeta^{+3}\,\zeta^{-2} + \frac{1}{2\sqrt{858}}\,\zeta^{+4}\,\zeta^{-3}\;,\\
\nonumber{[\mult \mult]}_{7}^{(-1)} &= \frac{1}{\sqrt{429}}\,\zeta^{-3}\,\zeta^{+3} + \sqrt{\frac{7}{143}}\,\zeta^{-2}\,\zeta^{+2} + \sqrt{\frac{35}{143}}\,\zeta^{-}\,\zeta^{+} + 5\sqrt{\frac{7}{429}}\,\zeta^{0}\,\zeta^{0}\\\nonumber&\hspace*{1em} + \sqrt{\frac{35}{143}}\,\zeta^{+}\,\zeta^{-} + \sqrt{\frac{7}{143}}\,\zeta^{+2}\,\zeta^{-2} + \frac{1}{\sqrt{429}}\,\zeta^{+3}\,\zeta^{-3}\;,\\
\nonumber{[\mult \mult]}_{7}^{(-2)} &= \sqrt{\frac{3}{286}}\,\zeta^{-3}\,\zeta^{+2} + \sqrt{\frac{35}{286}}\,\zeta^{-2}\,\zeta^{+} + \sqrt{\frac{105}{286}}\,\zeta^{-}\,\zeta^{0} + \sqrt{\frac{105}{286}}\,\zeta^{0}\,\zeta^{-}\\\nonumber&\hspace*{1em} + \sqrt{\frac{35}{286}}\,\zeta^{+}\,\zeta^{-2} + \sqrt{\frac{3}{286}}\,\zeta^{+2}\,\zeta^{-3}\;,\\
\nonumber{[\mult \mult]}_{7}^{(-3)} &= \sqrt{\frac{5}{143}}\,\zeta^{-3}\,\zeta^{+} + \sqrt{\frac{35}{143}}\,\zeta^{-2}\,\zeta^{0} + \sqrt{\frac{63}{143}}\,\zeta^{-}\,\zeta^{-} + \sqrt{\frac{35}{143}}\,\zeta^{0}\,\zeta^{-2} + \sqrt{\frac{5}{143}}\,\zeta^{+}\,\zeta^{-3}\;,\\
\nonumber{[\mult \mult]}_{7}^{(-4)} &= \frac{1}{2}\sqrt{\frac{5}{13}}\,\zeta^{-3}\,\zeta^{0} + \frac{1}{2}\sqrt{\frac{21}{13}}\,\zeta^{-2}\,\zeta^{-} + \frac{1}{2}\sqrt{\frac{21}{13}}\,\zeta^{-}\,\zeta^{-2} + \frac{1}{2}\sqrt{\frac{5}{13}}\,\zeta^{0}\,\zeta^{-3}\;,\\
\nonumber{[\mult \mult]}_{7}^{(-5)} &= \sqrt{\frac{3}{13}}\,\zeta^{-3}\,\zeta^{-} + \sqrt{\frac{7}{13}}\,\zeta^{-2}\,\zeta^{-2} + \sqrt{\frac{3}{13}}\,\zeta^{-}\,\zeta^{-3}\;,\\
\nonumber{[\mult \mult]}_{7}^{(-6)} &= \frac{1}{\sqrt{2}}\,\zeta^{-3}\,\zeta^{-2} + \frac{1}{\sqrt{2}}\,\zeta^{-2}\,\zeta^{-3}\;,\\
{[\mult \mult]}_{7}^{(-7)} &= \zeta^{-3}\,\zeta^{-3}\;.
\end{align}


\end{document}